\documentclass[pdflatex,sn-mathphys-num]{sn-jnl}

\usepackage{graphicx}%
\usepackage{multirow}%
\usepackage{amsmath,amssymb,amsfonts}%
\usepackage{amsthm}%
\usepackage{mathrsfs}%
\usepackage[title]{appendix}%
\usepackage{xcolor}%
\usepackage{textcomp}%
\usepackage{manyfoot}%
\usepackage{booktabs}%
\usepackage{algorithm}%
\usepackage{algorithmicx}%
\usepackage{algpseudocode}%
\usepackage{listings}%

\begin{document}

\title[Article Title]{Control of 2D plasmons in the topological insulator Bi$_2$Se$_3$ with highly crystalline C$_{60}$ overlayers}


\author[1]{\fnm{Mairi} \sur{McCauley}}\email{m.mccauley.1@research.gla.ac.uk}

\author[2]{\fnm{Lida} \sur{Ansari}}\email{lida.ansari@tyndall.ie}

\author[2]{\fnm{Farzan} \sur{Gity}}\email{farzan.gity@tyndall.ie}

\author[3]{\fnm{Matt} \sur{Rogers}}\email{m.d.rogers@leeds.ac.uk}

\author[3]{\fnm{Joel} \sur{Burton}}\email{py18j2b@leeds.ac.uk}

\author[3]{\fnm{Satoshi} \sur{Sasaki}}\email{S.Sasaki@leeds.ac.uk}

\author[3,4]{\fnm{Quentin} \sur{Ramasse}}\email{qmramasse@superstem.org}

\author[3]{\fnm{Craig} \sur{Knox}}\email{c.s.knox@leeds.ac.uk}

\author[2,5]{\fnm{Paul K.} \sur{Hurley}}\email{paul.hurley@tyndall.ie}

\author[1]{\fnm{Donald} \sur{MacLaren}}\email{donald.maclaren@glasgow.ac.uk}

\author*[3,6]{\fnm{Timothy} \sur{Moorsom}}\email{T.Moorsom@leeds.ac.uk}

\affil[1]{\orgdiv{SUPA, School of Physics and Astronomy}, \orgname{University of Glasgow}, \orgaddress{ \city{Glasgow}, \postcode{G12 8QQ}, \country{UK}}}

\affil[2]{\orgdiv{Micronano Electronics Group}, \orgname{Tyndall National Institute}, \orgaddress{ \city{Cork}, \postcode{T12R5CP}, \country{Republic of Ireland}}}

\affil[3]{\orgdiv{School of Physics and Astronomy}, \orgname{University of Leeds}, \orgaddress{ \city{Leeds}, \postcode{LS2 9JT}, \country{UK}}}

\affil[4]{\orgdiv{SuperSTEM}, \orgname{SciTech Daresbury Science and Innovation Campus}, \orgaddress{\street{Kedwick Lane}, \city{Daresbury}, \postcode{WA4 4AD}, \country{UK}}}

\affil[5]{\orgdiv{School of Chemistry}, \orgname{University College Cork}, \orgaddress{ \city{Cork},  \country{Ireland}}}

\affil[6]{\orgdiv{School of Chemical and Process Engineering}, \orgname{University of Leeds}, \orgaddress{ \city{Leeds}, \postcode{LS2 9JT}, \country{UK}}}


\abstract{Topological Insulators (TIs) present an interesting materials platform for nanoscale, high frequency devices because they support high mobility, low scattering electronic transport within confined surface states. However, a robust methodology to control the properties of surface plasmons in TIs has yet to be developed. We propose that charge transfer between Bi$_2$Se$_3$ and crystalline C$_{60}$ films may provide tunable control of the two-dimensional plasmons in Bi$_2$Se$_3$. We have grown heterostructures of Bi$_2$Se$_3$/C$_{60}$ with exceptional crystallinity. Electron energy loss spectroscopy (EELS) reveals significant hybridisation of $\pi$ states at the interface, despite the expectation for only weak van der Waals interactions, including quenching of 2D plasmons. Momentum-resolved EELS measurements are used to probe the plasmon dispersion, with Density Functional Theory predictions providing an interpretation of results based on interfacial charge dipoles. Our measurements suggest a robust methodology for tuneable TI interfaces that can be engineered for plasmonic applications in computing, communications and sensing.}




\maketitle

\section*{Introduction}\label{sec1}
A major barrier to the development of practical plasmonic devices is electronic scattering, which limits conduction efficiency, particularly in conventional metallic films. \cite{Khurgin2017} However, low dimensional systems including graphene offer high mobility and low scattering rates, making them attractive for the development of sensing and communications applications. \cite{Iranzo2018, Li2017, Grigorenko2012} Within these, topological insulators are especially interesting for operation at both THz and optical frequencies \cite{Lai2014, Stauber2014, Yin2017, DiPietro202} because of their topologically protected surface states (TSS). These arise from band inversion in the bulk \cite{Hasan2010} and can support Dirac Plasmon modes. \cite{DiPietro2013} Such collective excitations of two-dimensional (2D) Dirac states can exhibit suppressed scattering and spin pumping effects due to spin momentum locking, a feature that may enable spintronic-photonic integration. \cite{Stauber2013, Raghu2010}

Practical TI-based plasmonic devices will require a means of tunable control of 2D plasmons. One approach is to dope the surface with impurities to create a 2D electron gas (2DEG). \cite{Bianchi2010} For example, depositing Rb on freshly cleaved $Bi_{2}Se_{3}$ is observed to create a 2DEG with significant Rashba splitting. \cite{Bianchi2012} However, these dopants are extremely reactive, so the effect is unstable outside ultra high vacuum. A more robust alternative is the use of thin films of organic molecules and dyes, \cite{Kitazawa2020, Wang2014, Caputo2016}  although progress has been hampered by contradictory results, especially regarding Rashba coupling in these interfaces. The ultimate goal is to find a stable molecular thin film that supports reversible charge transfer to an underlying TI, so that it can be incorporated within a heterostructure where spin-plasmon behaviour can be tuned with the application of an external voltage \cite{Wei2022}.

Characterisation of TI heterostructures is complicated by the accessibility of the surface. To directly image the band-structure, techniques like ARPES, which show increased Rashba splitting in TI-molecular interfaces, are limited by penetration depth. \cite{whitcher2020correlated}\cite{Jakobs2015} Electrical or THz characterisation can struggle to distinguish interface and bulk effects except at very low temperatures. \cite{cai2018independence} STEM-EELS provides a useful probe of localised surface effects where interfacial electronic structure can be probed with sub-nm resolution. SPP modes in TI thin films are mostly in the THz regime. \cite{Jia2017} However, STEM-EELS can detect plasmon excitations above 1 eV which arise from interband excitations. In particular, the 2D $\pi$ plasmon mode observed in both topological interfaces \cite{Liou2013} and 2D materials such as graphene \cite{politano2014plasmon} provides a sensitive probe of 2D confined surface states. \cite{despoja2013two} In $Bi_2Se_3$, the $\pi$ plasmon mode has been observed in free-standing films and shown to correspond to the presence of a 2D surface state. \cite{Liou2013} Extending this methodology to TI heterostructures allows us to investigate how the 2D surface states, including the TSS, are modified by different overlayers and investigate materials for gate-tuneable topological plasmonics.


\section*{Results}\label{sec2}

\subsection*{Sample Preparation}\label{sec2b}

\begin{figure}[h]
    \centering
    \includegraphics[scale = 0.51]{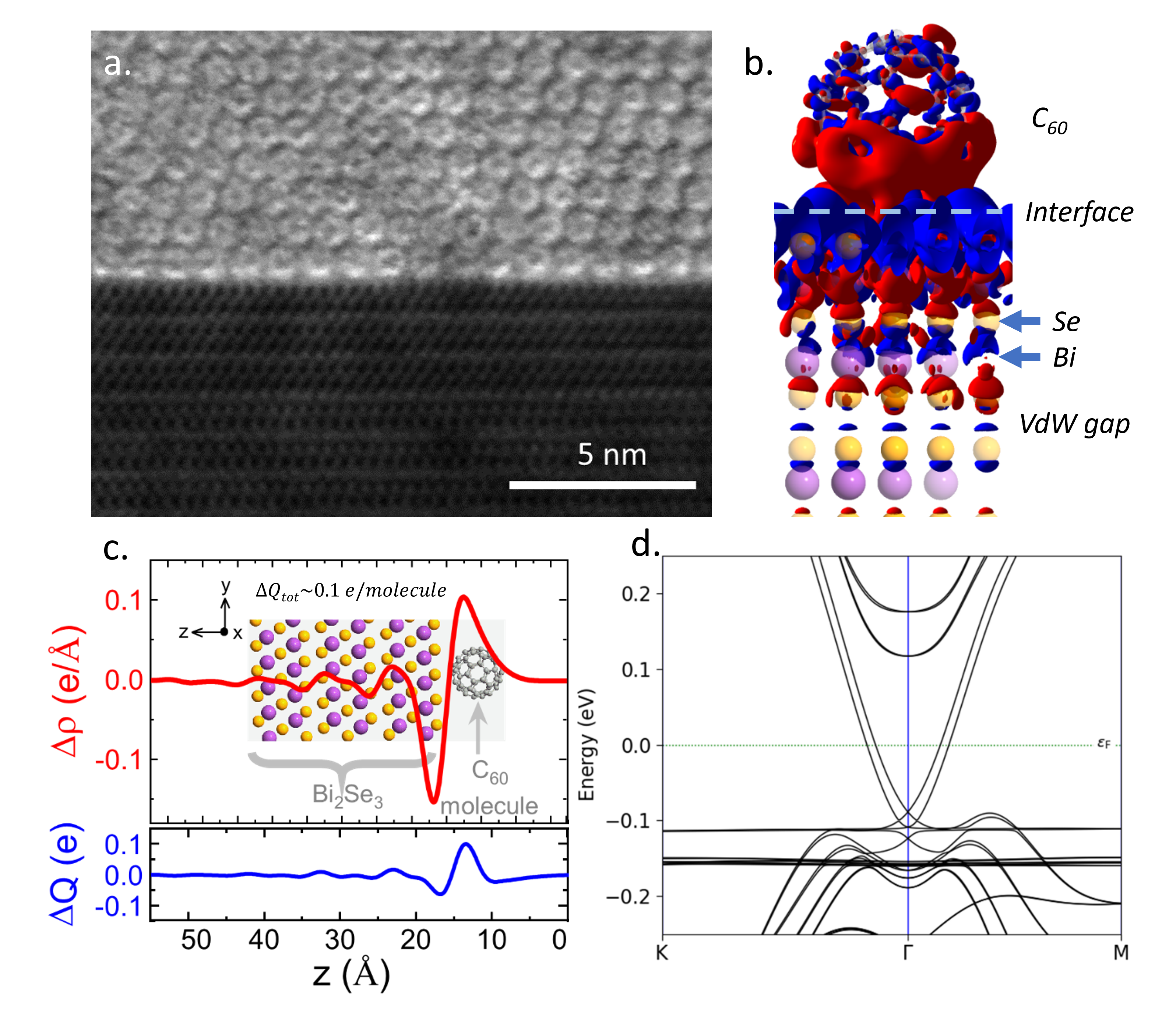}
    \caption{a. A cross-section bright field STEM image of $Bi_2Se_3$/$C_{60}$, showing an atomically sharp interface between the $Bi_2Se_3$ and a highly crystalline, ordered overlayer of $C_{60}$. Note that only the heavy Bi atoms are readily resolved in the $Bi_2Se_3$, with van der Waal gaps appearing as brighter horizontal bands b. Atomic structure of the $Bi_2Se_3/C_{60}$ supercell with charge density difference showing charge accumulation (red) and depletion (blue) around the molecular absorption site. The interface is marked with the dotted line. Se atoms are yellow and Bi atoms are purple. c. Change in charge density [top] and transferred charge [bottom] across the interface. d. Band structure of the surface state as predicted in DFT.}
    \label{fig:C60 STEM}
\end{figure}

C$_{60}$ films may form an ideal, stable molecular overlayer for the control of plasmonic transport in $Bi_{2}Se_{3}$ due to their high electron affinity and chemical robustness. A region of the interface between $Bi_{2}Se_{3}$ and continuous films of highly crystalline C$_{60}$ is shown in cross-sectional scanning transmission electron microscopy (STEM) in Figure \ref{fig:C60 STEM}a. We have found that $C_{60}$ forms highly ordered crystals on TI surfaces at very low deposition energy, producing atomically sharp interfaces that are free from structural defects and stable to degradation (see Methods). C$_{60}$ was chosen for its high electron affinity, resulting in significant charge transfer from the TI interface. \cite{Latzke2019} Density Functional Theory (DFT) simulations confirm the expectation for significant charge transfer from the uppermost layer of the TI to the molecular cage of the C$_{60}$, shown in figure \ref{fig:C60 STEM}b and c. The interaction between the two materials is also predicted to give rise to Rashba splitting of the surface state and the formation of flat bands, evident in the band-structure in figure \ref{fig:C60 STEM}d.

Bi$_2$Se$_3$/C$_{60}$ hybrid structures were grown in a multi-functional molecular beam epitaxy (MBE) system under continuous ultra high vacuum (UHV). 15 QL (quintuple layers), just over $15~$nm thickness, of Bi$_2$Se$_3$ were deposited onto a c-plane sapphire substrate using a self regulating growth method. Detailed electronic characterisation of material produced from this system and the growth methodology are outlined in previous works \cite{pistore2024terahertz} and in the Supplemental Information. C$_{60}$ was then deposited using a low-temperature Knudsen cell, to a thickness of $\sim 80~nm$. Cross-sections were extracted and thinned to $\sim 35~nm$ using a Focused Ion Beam (FIB), further details of which are provided in Methods. The films that we have grown show extraordinary crystallinity for a hybrid molecular film, to the extent that individual fullerene columns can easily be resolved. EELS measurements suggest the C$_{60}$ to be of high purity without detectable oxygen content. The films also have a remarkably low defect density for a van der Waals bonded molecular film. The Fourier Transform of a STEM image of the C$_{60}$ is dominated by sharp spots, indicative of high crystallinity. Geometric phase analysis, highlights a low density of stacking faults, running diagonally through the film with spacing of order 20 nm. These typically nucleate at step edges in the underlying Bi$_2$Se$_3$/C$_{60}$. This analysis is shown in Methods.


\begin{figure}[h]
    \centering
    \includegraphics[scale = 0.6]{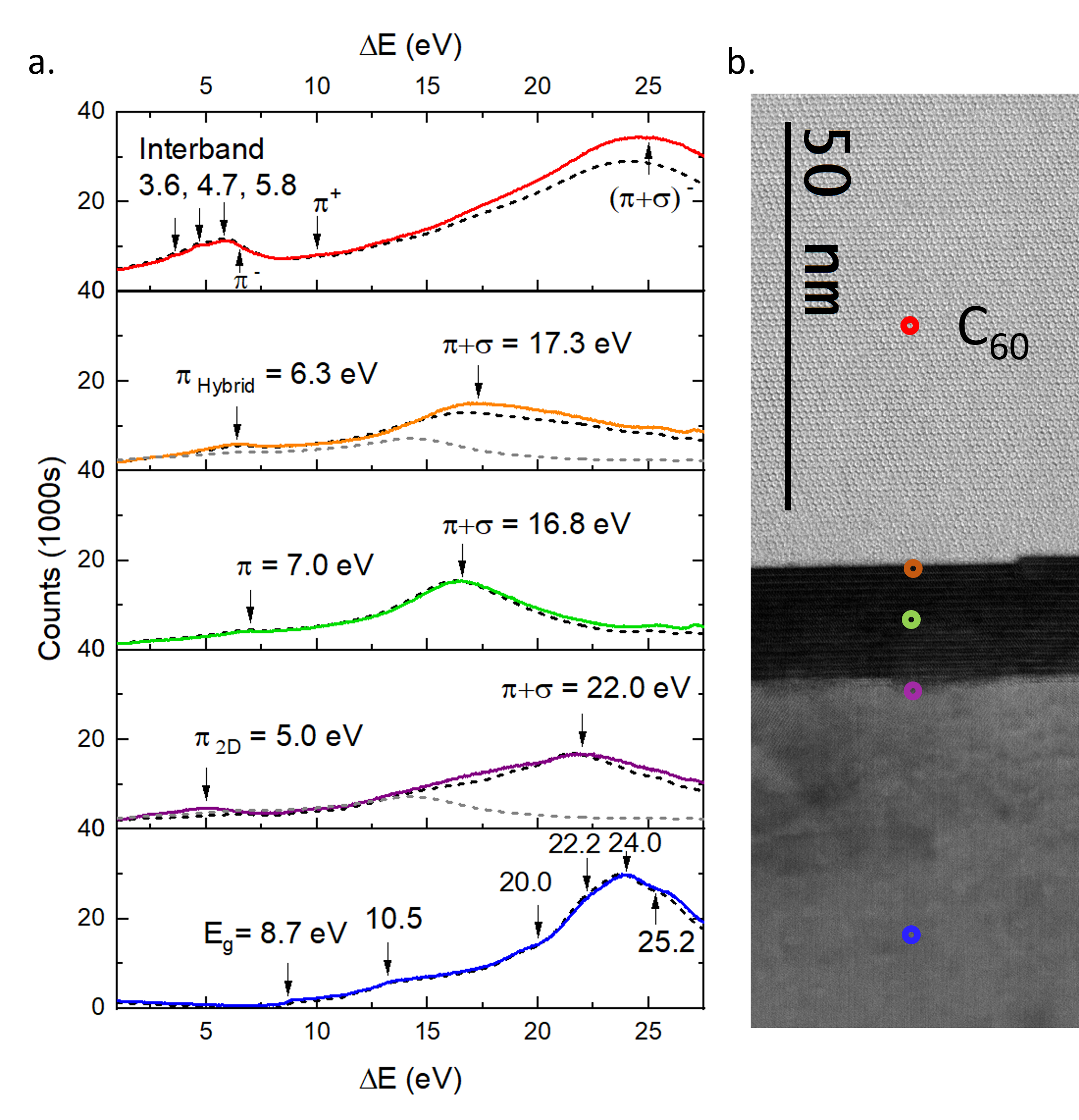}
    \caption{a) EELS spectra collected from $C_{60}$ (red), $C_{60}$/$Bi_2Se_3$ interface (orange), bulk $Bi_2Se_3$ (green), $Bi_2Se_3$/$Al_2O_3$ interface (purple) and bulk $Al_2O_3$ (blue), with key features marked. Analytical models of the EELS spectra, derived from empirical dielectric functions, are marked as dotted lines. These models are described in more detail in the Methods. b) Cross-sectional STEM image of the sample lamella, showing approximate positions for the STEM probe for each spectrum.  }
    \label{fig:spectra}
\end{figure}



\subsection*{EELS Spectrum Imaging}\label{sec2b}

To study the sample's plasmonic structure, we mapped the EELS spectra below 28 eV across a cross-section of the sample. Figure \ref{fig:spectra}a shows spectra collected from the positions indicated in colour in the STEM image of Figure \ref{fig:spectra}b. In the $C_{60}$ film, figure \ref{fig:spectra}a [red], a cluster of three peaks in the 4-6 eV energy range is known to correspond to interband transitions between the highest occupied and lowest unoccupied molecular orbitals (HOMO and LUMO, respectively): specifically, from the HOMO to LUMO+1 at 3.6 eV, HOMO to LUMO+2 at 4.7 eV and HOMO-1 to LUMO at 5.8 eV. \cite{Ostling1993} As observed previously \cite{Barton1991, Ostling1993}, the features at 6.8 eV, 10 eV and 25 eV correspond to the $\pi^-$, $\pi^+$ and $(\pi + \sigma)^-$ plasmonic modes of $C_{60}$, respectively. These are excitations of the induced charge on the shell of the molecule. The $\pi$ and $\sigma$ labels denote the orbitals excited \cite{Barton1991} while the $-/+$ labels indicate the anti-symmetric/symmetric modes, where charge oscillations on the inner and outer surfaces of the $C_{60}$ cage are either out of phase or in phase. \cite{Ostling1993, BOLOGNESI2012, HANSEN1991} The $\pi^-$ plasmon at 6.8 eV overlaps with the LUMO interband transitions. \cite{GOROKHOV1996} Other features, at 13, 14.6 and 17 eV, are generally attributed to different ionization states of the fullerene. \cite{BOLOGNESI2012}

\begin{figure}[h]
    \centering
    \includegraphics[scale = 0.35]{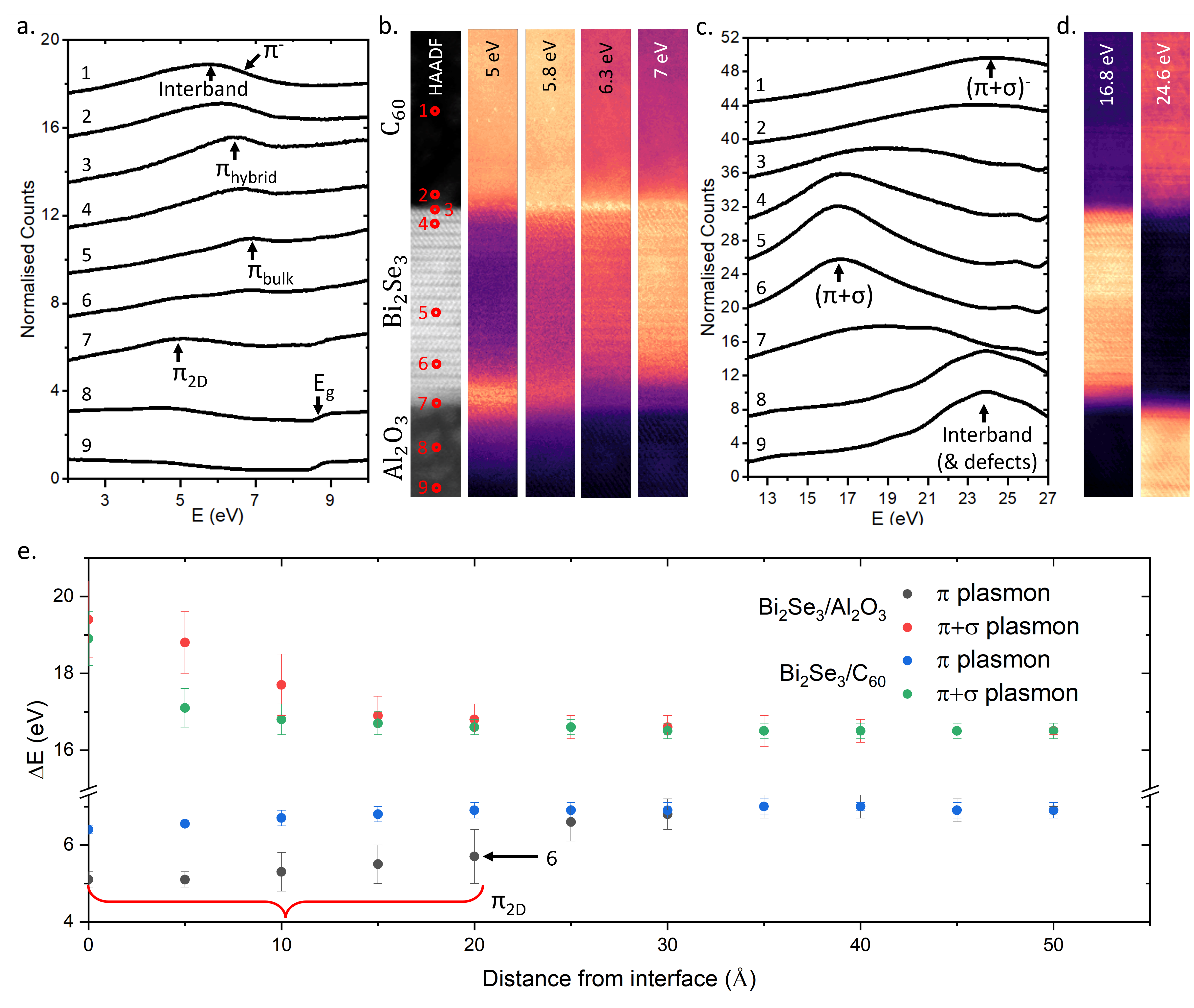}
    \caption{EELS spectrum images collected across the two interfaces of $Al_2O_3$/$Bi_2Se_3$/$C_{60}$. a) EELS spectra collected from the region of the $\pi$ plasmon with the STEM probe at the positions marked 1-9 in b. The important features in each spectrum are indicated. b) HAADF and spectral maps of a 3x30 nm region of the sample. Each spectral map shows the spectral density over the given energy window, normalised to the zero loss. The spectrum between 4.8 - 5.2 eV shows the peak height relative to the background intensity. c) EELS spectra in the region of the $\pi + \sigma$ plasmons located with the STEM probe at the positions marked 1-9 in b. d) Spectral density across the same region as b. within the specified energy ranges. e) Energy of the $\pi$ and $\pi+\sigma$ plasmon modes in $Bi_2Se_3$ with distance from the interfaces. The data point at the position marked 6 in b. is indicated.}
    \label{fig:Interface}
\end{figure}


In the bulk $Bi_2Se_3$, two excitations are observed, at 7 eV and 16.8 eV. In previous work, these are ascribed to $\pi$ and $\pi+\sigma$ plasmon modes respectively, relating to plasmonic excitations of $\pi$ and $\sigma$ bonds in the Se layers. \cite{Liou2013} For the sapphire substrate, the band gap at 8.7 eV is clearly evident in figure \ref{fig:spectra}a [blue], followed by interband transitions at 10.5 eV and 13.2 eV. These are in agreement with literature, while transitions at 20.0 eV, 22.2 eV, 24 eV and 25 eV have all been observed in $Al_2O_3$ crystals with small numbers of defects that here we attribute to the focused ion beam sample preparation. \cite{GILLET1992, GILLET1993} 

It is clear that the two interfaces have different spectral characteristics to those of bulk materials. Typically, an interfacial plasmon will depend on the dielectric functions of the surrounding materials, and the sample's response can be considered to arise from a coupling of excitations in both materials due to the transmitted electron beam its image charge. In order to distinguish the bulk and interfacial contributions, each spectrum is compared with an analytical model, shown by dotted lines in figure \ref{fig:spectra}a. The model employs a retarded-field approach described previously (see also Methods) \cite{Maclean2001,Konecna2018} and uses empirically-determined dielectric functions. Since these spectra were calculated using bulk dielectric functions, we can isolate those excitations that are distinctive of surface and interfacial electronic structures not present in the bulk. In the spectral region of interest here, around the energy of the $\pi$ plasmon below 10~eV, significant differences between model and experiment are observed at both interfaces. At the interface between the $C_{60}$ and $Bi_2Se_3$, there is a single peak at 6.3 eV. This excitation lies half way between the $C_{60}$ $\pi^-$ plasmon and the $Bi_2Se_3$ bulk $\pi$ plasmon, thus we label it a hybrid $\pi$ - plasmon. At the interface between the $Bi_2Se_3$ and $Al_2O_3$, a distinct 2D $\pi$ plasmon mode is evident, first appearing within 2QL of the interface. This excitation has also been observed at $Bi_2Se_3$/vaccuum interfaces \cite{Liou2013}, and can be excited in an aloof position well into the substrate, appearing in the $Al_2O_3$ band-gap more than 4 nm from the interface.  \cite{KRIVANEK2019}

Figure \ref{fig:Interface} shows in more detail the region of the $\pi$ plasmon across a 3x30 nm region of the sample. The spectrum labels 1-9 correspond to the points marked on the HAADF image in figure \ref{fig:Interface} b, which shows intensity maps across the same sample region over various energy ranges. All maps are normalised to the zero loss intensity and the $\pi$ plasmon contribution is deconvoluted from the volume plasmons by fitting each spectrum with a pseudo-Voigt function and plotting the residuals. This removes artefacts in the spectral images caused by the tail of the volume plasmons increasing the apparent intensity of the $\pi$ plasmon modes at the interfaces, detailed in Supplemental Information. 

At 5 eV, the $Bi_2Se_3$ 2D $\pi$ - plasmon mode is shown to be strongest at the $Bi_2Se_3/Al_2O_3$ interface and has non-zero amplitude in the band-gap of the sapphire. This mode decays exponentially into the substrate, indicating aloof excitation of the surface. The 2D mode also appears well into the $Bi_2Se_3$, first emerging in spectrum 6, 2QL from the interface. This raises interesting questions about the physical confinement of surface states in TIs. The $C_{60}$ appears bright at this energy due to the tail of the broad interband transition peak in $C_{60}$ highlighted in figure \ref{fig:spectra} a, which extends well below 5 eV. There is, however, no evidence of a 2D plasmon at the $Bi_2Se_3$/$C_{60}$ interface, indicating this behaviour is quenched by the $C_{60}$. At 5.8 eV, the peak of the interband transition in $C_{60}$ can be isolated, which are confined to the $C_{60}$ film, overlapping somewhat with the 2D $\pi$ plasmon. 

6.3 eV corresponds to the expected energy of what we have labelled the hybrid $\pi$ - plasmon mode. This range overlaps with both the bulk $\pi$ - plasmon mode in $Bi_2Se_3$ and the interband excitations in $C_{60}$, meaning there is a residual non-zero intensity in both bulk regions. The spectral intensity peaks at the $Bi_2Se_3/C_{60}$ interface, which can also clearly be seen in spectrum 3 in figure \ref{fig:Interface}a showing that this is a surface effect. Interestingly, the contact points between individual fullerenes and the $Bi_2Se_3$ surface can be easily discerned as bright spots in the spectral map. This implies the 6.3 eV hybrid plasmon is not a 2D excitation, as might be expected, but an oscillation of the local charge dipole predicted in DFT. The 7 eV  map highlights the bulk $\pi$ - plasmon mode in $Bi_2Se_3$, which does not show this surface enhancement at either interface.

The $\pi+\sigma$ volume plasmons in $Bi_2Se_3$ and $C_{60}$ are shown in figure \ref{fig:Interface} c and d, and show no surface amplification. The high surface intensity of the hybrid $\pi$ - plasmon can be explained by Surface Enhanced EELS (SEELS). EELS is often more sensitive to molecular excitations at conducting interfaces due to resonant interactions between interband transition states and plasmon excitations in the surface. \cite{Konecna2018_SEELS} 


\begin{figure}[h]
    \centering
    \includegraphics[scale = 0.6]{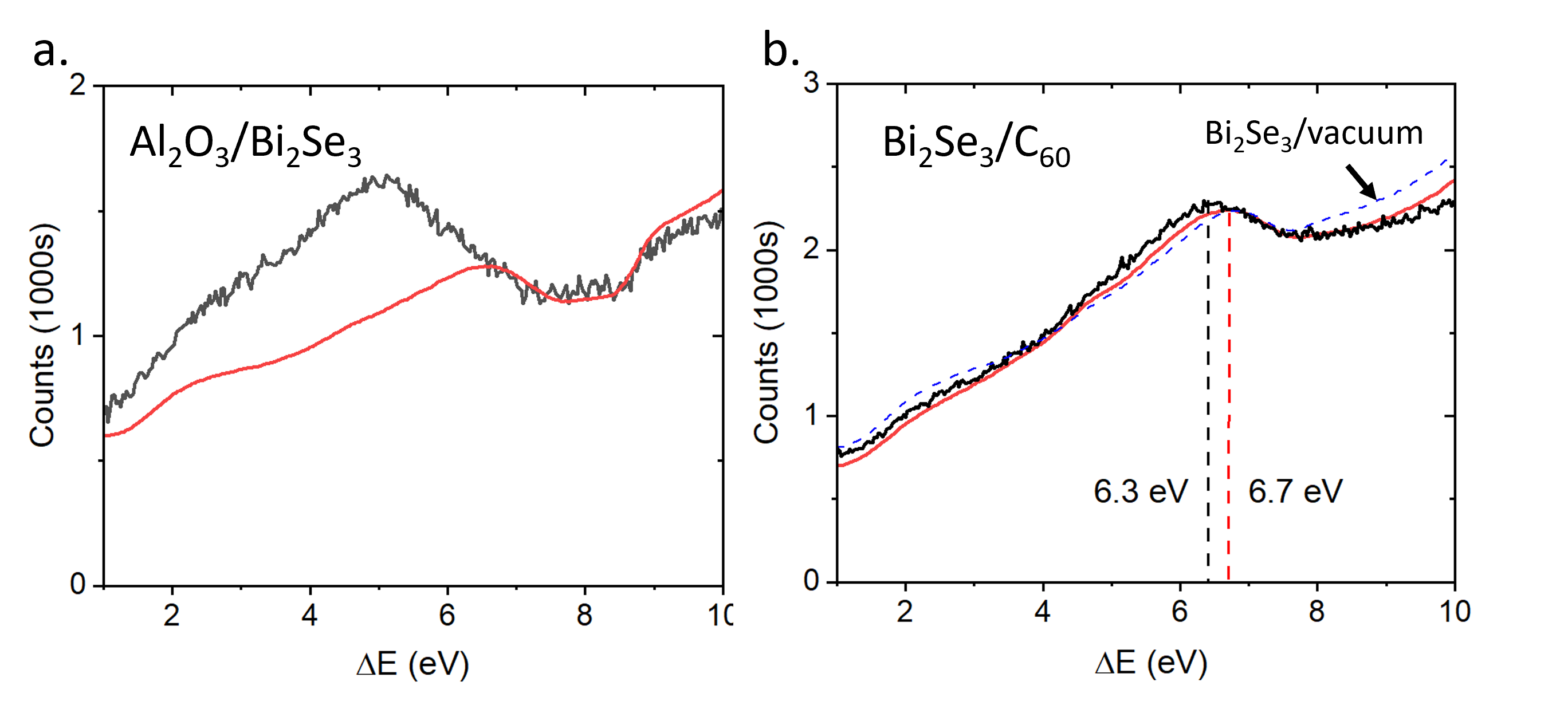}
    \caption{Comparisons of the EELS model (red) and $\pi$-plasmon modes (black) at (a) the $Bi_2Se_3$/$C_{60}$ interface and (b) the $Al_2O_3$/$Bi_2Se_3$ interface. The peak in panel a is not well reproduced by the model while panel b shows a slight peak-shift with respect to the model. Both aspects indicate suggest that the observed surface features are not accounted for by red-shifted bulk plasmons. The blue dotted line in b shows an analytical model for a $Bi_2Se_3$ interface with vacuum.}
    \label{fig:Simulations}
\end{figure}

The shift of the $\pi$ and $\pi + \sigma$ plasmon modes in $Bi_2Se_3$ are shown in figure \ref{fig:Interface}e. At the $Bi_2Se_3$/$C_{60}$ interface, the total redshift of the $\pi$ - plasmon mode is just 0.7 eV, while the 2D $\pi$ - plasmon mode seen at the $Bi_2Se_3$/$Al_2O_3$ interface is 2 eV below the bulk, identical to vacuum interfaces \cite{Liou2013}. This suggests that it is not simply the difference in dielectric constant that alters the the $Bi_2Se_3$/$C_{60}$ interface, and the $C_{60}$ film has distinct effects on the TI surface states, which is supported by previous ARPES studies. \cite{Jakobs2015}

A comparison of the $\pi$-plasmon region to the analytical model shows that the 2D $\pi$ plasmon is not predicted by bulk dielectric functions as expected, Figure \ref{fig:Simulations}a. The model of the $Bi_2Se_3$/$C_{60}$ interface shows much closer agreement with experiment, as shown in Figure \ref{fig:Simulations}b. The difference in the predicted redshift of the $\pi$ plasmon at the $Bi_2Se_3$/$C_{60}$ interface is likely because the model does not account for hybridisation of the $C_{60}$ molecular orbitals with the TI surface $\pi$ electrons. As has been previously reported in graphene, \cite{shin2011control} $\pi$ surface plasmons are very sensitive to charge doping and hybridisation.

\subsection*{Hybrid Plasmons}\label{sec3}
\begin{figure}[h]
    \centering
    \includegraphics[scale = 0.65]{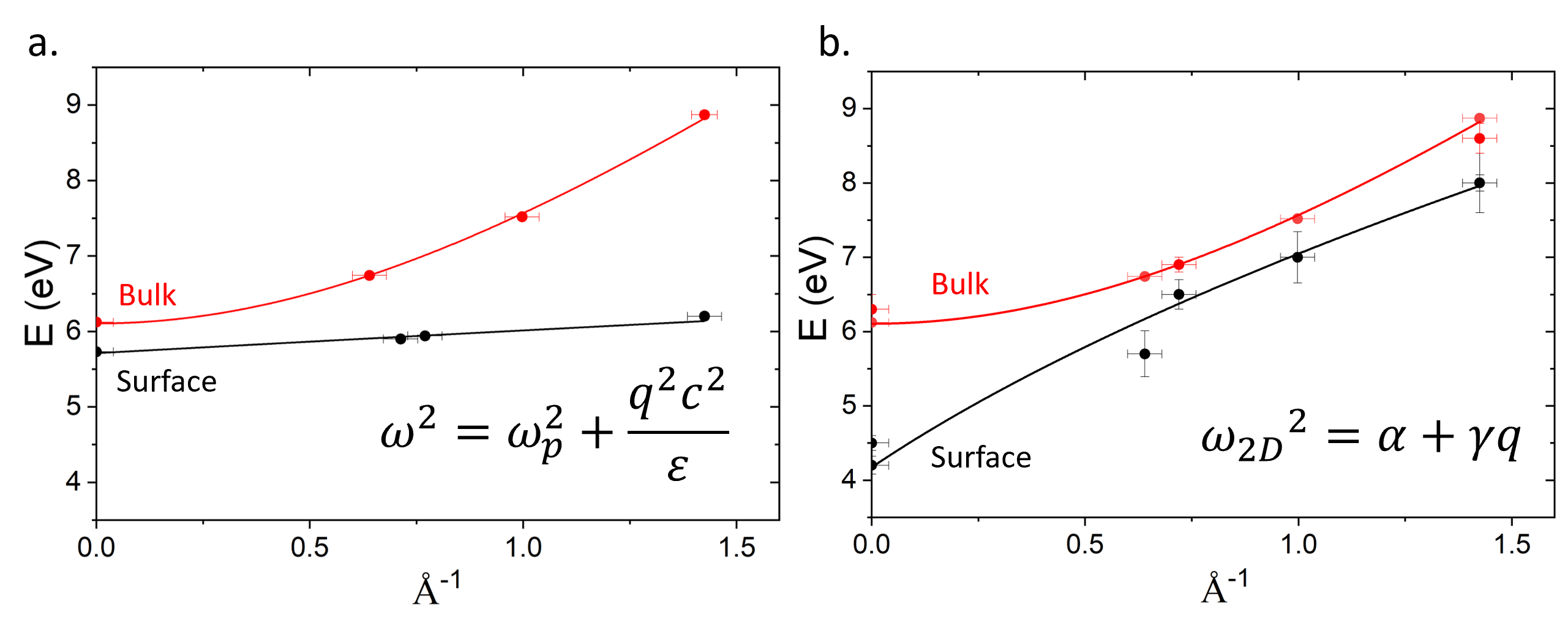}
    \caption{a) Dispersion for the bulk (red) and interface (black) $\pi$-plasmon mode obtained from QEELS spectra at the TI/C60 interface. The interface spectrum shows very weak dispersion. b) Dispersion for the bulk (red) and interface (black) $\pi$-plasmon mode obtained from QEELS spectra at the $Al_2 O_3$/TI interface. Here, the surface obeys the $\sqrt{q}$ dispersion characteristic of a 2D plasmon excited via an interband transition. \cite{NAGASHIMA1992}} 
    \label{fig:qEELSdata}
\end{figure}

A deeper understanding of the effects of the $C_{60}$ layer at the TI interfaces can be obtained by analysing changes in dispersion. Momentum-resolved EELS (QEELS) spectra from $Bi_2Se_3$ are shown in figure \ref{fig:qEELSdata}. Spectra were recorded with a momentum resolution of $\pm 0.04$ \AA$^{-1}$ over the range 0 - 1.43 \AA$^{-1}$ and in the $\Gamma - M$ direction. In QEELS, all peaks appear shifted down in energy with respect to the unresolved EELS spectrum, with a difference in the energy of the q = 0 peak of $\approx$ 300 meV. This is because integration over the larger, 22 mrad aperture used for STEM-EELS is weighted towards higher Q in comparison to the 2.4 mrad aperture used for QEELS (the correction is discussed further in the Supplementary Information). 

Plasmon peaks were fitted with pseudo-Voigt functions at each momentum with peak energy and peak width as free parameters (details of the fitting approach can be found in the Supplementary Information.) In bulk $C_{60}$, the three interband transitions were not observed to disperse with momentum, as expected for localised excitations. However, the hybrid $\pi$ - plasmon also shows almost zero dispersion at the $Bi_2Se_3$/$C_{60}$ interface. At the interface between $Bi_2Se_3$ and $Al_2O_3$, the $\pi$-plasmon peak followed a 2D plasmon dispersion with a $q^{1/2}$ dependence, as shown in figure \ref{fig:qEELSdata}b. \cite{Liou2015, Stern1967}

Liou et al. proposed the dispersion of 2D $\pi$-plasmons in $Bi_2Se_3$ followed:
\begin{equation}
\label{eq:Liou_dispersion}
    \omega_p(q)^2 = \beta + \gamma q, 
\end{equation}
where $\beta$ is the single particle oscillator strength and $\gamma = 2\pi n_{2D}e^2/m\varepsilon$. This is the same behaviour observed in free standing graphene. \cite{Liou2015, Stern1967, NAGASHIMA1992} Upon fitting the dispersion data for the plasmon confined to the $Bi_2Se_3$/$Al_2O_3$ interface with equation \ref{eq:Liou_dispersion}, $\beta$ was 4.14$\pm$0.02 eV and the 2D electron density was obtained from $\gamma$. The effective mass of $\pi$ electrons excited into the parabolic band in the $\Gamma - M$ direction was 1.05 $m_0$ obtained from DFT. The permittivity was estimated from Kramers-Kr{\"o}nig analysis of bulk data as $\epsilon_1 = 2.35$, giving the 2D electron density of the $\pi$-plasmon as $1.9 \pm 0.1\times10^{14}$cm$^{-2}$. This value is close to the estimated number of $\pi$ bonded electrons in the surface Se layer of $Bi_2Se_3$ which was calculated to be $\approx 7\times10^{14}$cm$^{-2}$ while the total number of $\pi + \sigma$ electrons in the surface Se layer was found from DFT to be $2\times10^{15}$cm$^{-2}$, meaning roughly one in three $\pi$ bonded electrons in the surface Se layer contribute to the $\pi$ plasmon in this geometry. Details of this calculation can be found in the Supplemental Information.

Fitting the $Bi_2Se_3$/$C_{60}$ interface with equation \ref{eq:Liou_dispersion}, $\beta$ was measured to be 5.71$\pm$0.01 eV. The $\pi$ plasmon at this interface showed almost no dispersion, with the surface plasmon peak overlapping the expected $C_{60}$ interband transition up to the edge of the first BZ. Using a relative permittivity of $C_{60}$ at 6.3 eV of $\approx$ 1, \cite{kataura1997dielectric} the 2D electron density found from equation \ref{eq:Liou_dispersion} was $2.14 \pm 0.01\times10^{13} 
 cm^{-2}$, an order of magnitude lower than that of the TI/insulator interface. However, it is more likely this is a non-dispersive excitation of the hybrid $\pi$ electrons at the interface, involving an oscillation of the surface charge distribution shown in Figure \ref{fig:C60 STEM}b. This evidences that the hybrid $\pi$ plasmon is fundamentally distinct from the 2D $\pi$ plasmon reported at TI-vacuum interfaces, which is strongly quenched at the molecular interface. This indicates the significant changes the $C_{60}$ layer has induced in the surface $\pi$ states.

DFT simulations allow us to analyze the ways in which $C_{60}$ modifies the $Bi_2 Se_3$ surface. We find that the electron affinity of the $C_{60}$ molecule leads to the formation of an interfacial dipole, with a total of 0.1 e transferred to each molecule, which was demonstrated in figure \ref{fig:C60 STEM}b and c. The formation of this dipole explains the suppression of the 2D $\pi$ - plasmon and the emergence of the hybrid plasmon at 6.3 eV, figure \ref{fig:Interface} a and b. The behaviour of the $Bi_2 Se_3/C_{60}$ interface is not well explained by the formation of a 2DEG due to band-bending, which is expected in many TI heterostructures and interfaces. \cite{Bianchi2010} Though DFT predicts band bending at this interface, it is $\approx$ 100 meV, which is insufficient to populate the parabolic bands at the surface as shown in the relaxed band-structure in Figure \ref{fig:C60 STEM} d, implying the observed changes cannot simply be due to the formation of a 2DEG. Hybridisation between molecular orbitals and the surface states in the DFT model predicts flat bands, an indicator of very high density of states and even a possible 2D superconducting state arising from hybridised molecular orbitals. \cite{tang2014strain} This prediction, as well as the resonant excitations at the interface warrant further investigation. 

DFT also predicts Rashba splitting and a complex spin texture in the hybrid surface. In literature, ARPES studies of $Bi_2Se_3$ interfaces doped with single $C_{60}$ molecules did not show Rashba splitting, requiring 2D molecules such as $H_2Pc$ to see this effect. \cite{Jakobs2015} However, other studies of continuous molecular films showed Rashba splitting even for much weaker surface coupling. \cite{Kitazawa2020} Future work may confirm whether this Rashba splitting prediction is correct, in which case these highly crystalline layers provide a powerful means to manipulate surface spin-texture. Since the surface $\pi$ plasmon in 2D systems is extremely sensitive to changes in surface band structure, \cite{shin2011control}, the strong effect of $C_{60}$ doping on the $\pi$ plasmon dispersion in figure \ref{fig:qEELSdata}a is notable, since it implies that molecular layers can have significant effects on the surface state through charge transfer and hybridisation of the $\pi$ electron states, which may be electrically tunable. \cite{cinchetti2017activating} To explore the degree to which this surface can be electrically tuned, more work is needed, particularly utilising doped fullerenes such as Li@C60.

\section*{Discussion}

In summary, the engineering of interfaces in TI heterostructures is vital to the development of real world applications for topological insulators. Fullerenes present an interesting test system to explore the kinds of modifications that are possible at such engineered interfaces. We have demonstrated the deposition of highly crystalline films of $Bi_2Se_3$/$C_{60}$ in which individual fullerenes can be imaged at the interface. We have shown that this interface has very different behaviour to the 2D $\pi$-plasmon measured at the TI/insulator or TI/vacuum interfaces, evidenced by a non-dispersive hybrid excitation confined to the interface, and quenching of the 2D $\pi$ surface plasmon. Using DFT, we have simulated this interface to show how charge transfer gives rise to these emergent effects, which may be tuneable either via gating or doping. We also predict that this interface may host exotic topological surface textures, including Rashba-split parabolic bands and flat bands that may indicate a 2D superconducting state. Together, this analysis demonstrates that molecules can be used to control the surface properties of TI thin films. Future work should focus on exploring the effects of doping and gate control on this interface in order to actively modify the interfacial dipole and tune the properties of the surface. 

\backmatter

\section*{Methods}\label{sec11}

\subsection*{Sample Preparation}

$Bi_{2}Se_{3}/C_{60}$ hybrid structures were grown using a multi-functional molecular beam epitaxy (MBE) system, the Royce Deposition System, located at the University of Leeds. This system comprises linked chalchogenide MBE and organic MBE chambers, with a UHV transfer system that allows multi-step depositions to be performed under continuous UHV, eliminating the need for wet processing of organics, plasma cleaning or etching processes. TIs were deposited under a pressure of $10^{-9} mbar$ onto a c-plane sapphire substrate held at $230-240 ^\circ C$. Growth was monitored using a RHEED system. The film grew progressively as a series of van der Waals coupled QL. Growth was solely determined by bismuth flux, with the Se flux at least 20 times higher to insure no chalcogenide vacancies form. This growth is self-regulating, as the sticking coefficient of the top surface changes as the QL builds up. The THz behaviour and structural qualities of these films are outlined in recently published work. \cite{Knox2022} Once the film was grown and cooled, it was transferred under $10^{-10} mbar$ into an organics MBE. Here, the organic film was grown from a low temperature evaporation cell with the substrate held at $\sim 100 ^\circ C$. The growth of the molecular films was monitored by a quartz balance. Pressure and trace gasses are monitored live by a ThorLabs RGA system, with total pressure not exceeding $10^{-7} mbar$ and $O_2$ pressure not exceeding $10^{-10} mbar$.

 \begin{figure}[h]
    \centering
    \includegraphics[width = 0.7\columnwidth]{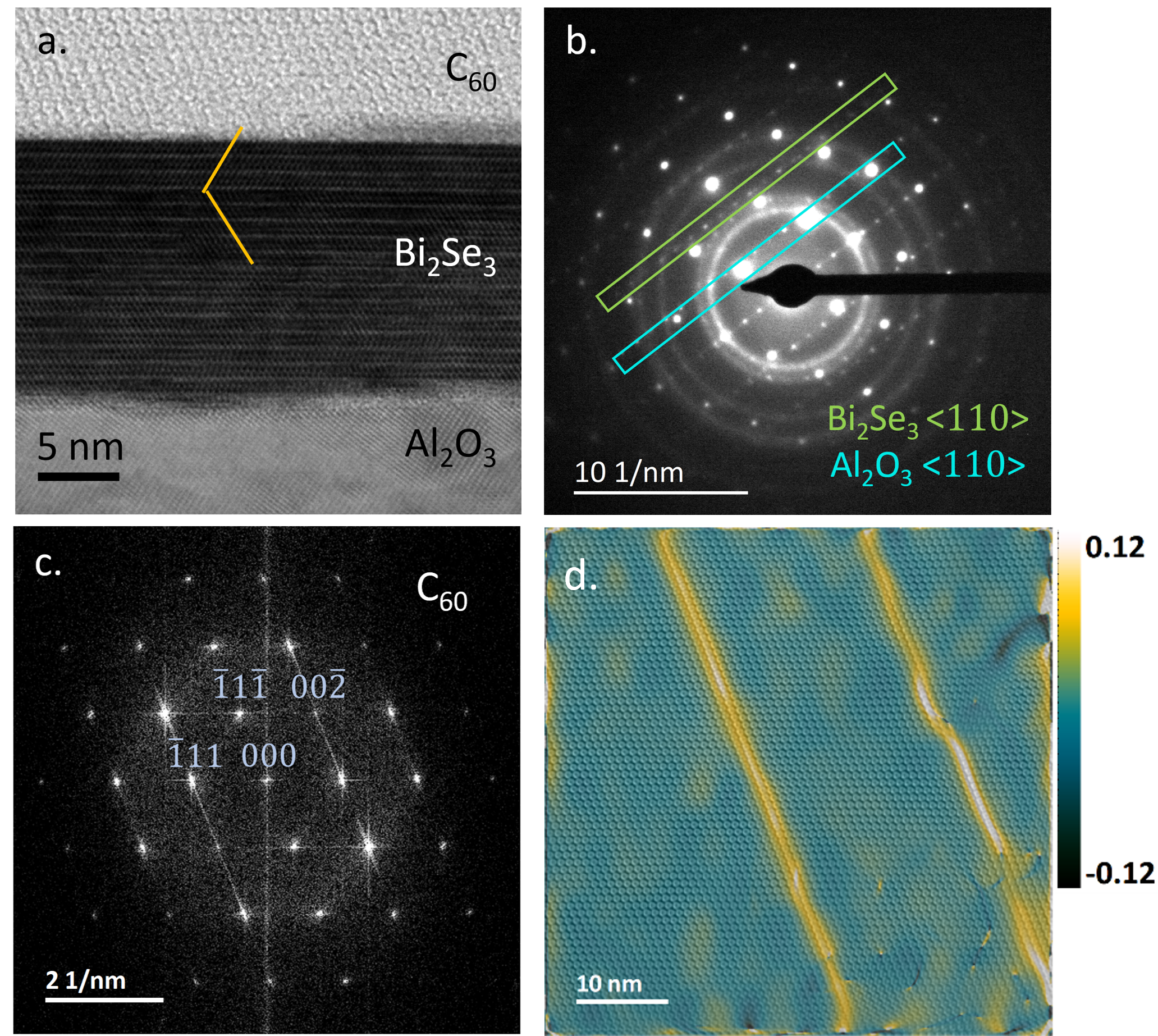}
    \caption{a. BF STEM image of the heterostructure. There is a twin plane 3QL from the surface. This common defect is often caused by surface strain and can be suppressed through substrate doping. \cite{Tarakina2014} b. TEM diffraction pattern and indexing of the $Bi_23Se_3$ film and substrate. c. Fourier transform of a STEM image with indexing of the $C_{60}$ film. d. GPA showing strain in the C$_{60}$ film, highlighting stacking faults that run parallel from the surface with spacing of 20 nm.} 
    \label{fig:structure}
\end{figure}

 Electron transparent lamella of the samples were prepared by focused ion beam (FIB) lift-out techniques using a dual beam electron-beam/FIB instrument, a Thermo Fisher Helios Xe Plasma FIB. A 30 kV xenon beam was used to mill into the bulk with currents 6.7 nA and 1.8 nA to extract a section which was subsequently thinned to a 35 nm thickness using a current of 74 pA and polished with a 5 kV 47pA ion beam. Initial TEM characterisation was carried out at 200kV to confirm the thickness and crystallinity of the samples before imaging in STEM. No direct epitaxial ordering was observed or expected between the film and substrate, as the surface of the sapphire is passivated and will form only weak van der Waals bonds with the thin film. The QLs grew with the c-axis parallel to the growth direction, such that the terminating surface of the last QL is exactly parallel to the substrate. The c-axis lattice constant is 28.6 \AA  while the quintuple layer thickness is 9.5 \AA. Samples vary between 12- 15 QLs thick across their surface, with single QL terraces forming at intervals of 20 - 50 nm.


\subsection*{EELS Model}

To model the interfacial EELS spectra, each material's response to the electric field of the STEM electron beam was first measured and used to derive a bulk dielectric function through a Kramers-Kr{\"o}nig analysis. \cite{Egerton2011} EELS spectra, $\Gamma$, between materials with dielectric functions $\varepsilon_1$ and $\varepsilon_2$ at a distance $b$ from the interface in material 1 were then calculated using:
\begin{multline}
\label{analytical_calc}
    \Gamma = \frac{e^2}{2\pi^2\varepsilon_0\hslash v_e^2} \int_0 ^{q_y^c} \Im \left\{  -\frac{1-\varepsilon_1\left(\frac{v_e}{c}\right)^2}{\alpha_1\varepsilon_1} \right. \\ 
    \left. + \frac{e^{-2\alpha_1b}}{\varepsilon_1\alpha_1(\alpha_1+\alpha_2)}\left[ \frac{2\alpha_1^2(\varepsilon_2-\varepsilon_1)}{\varepsilon_1\alpha_2+\varepsilon_2\alpha_1}+(\alpha_2-\alpha_1)\left(1-\varepsilon_1\left(\frac{v_e}{c}\right)^2\right)\right] \right\} dq_y
\end{multline}
where $q_y^c$ is a cutoff momentum defined by the spectrometer aperture, the electron velocity is $v_e$, $\alpha_i^2 = ((\omega/v)^2 + q_y^2) - \epsilon_i\omega^2/c^2 $ and fundamental constants have their usual symbols. \cite{Konecna2018,Maclean2001,GarciaMolina1985} The cross-section thickness, $t$, was determined using a known mean free path, $\lambda$ and measuring $t/\lambda$ directly by EELS. \cite{Egerton2011}

 \subsection*{EELS Methods}
 STEM-EELS measurements were carried out using the SuperSTEM3 instrument, a Nion UltraSTEM 100MC HERMES, incorporating a probe corrector and monochromator capable of producing a beam energy spread of 5 meV. Spectra were recorded with a beam energy of 100 keV and using Nion IRIS spectrometer, employing a 30 mrad convergence angle and a 22 mrad collection angle.

 \begin{figure}[h]
    \centering
    \includegraphics[scale = 0.8]{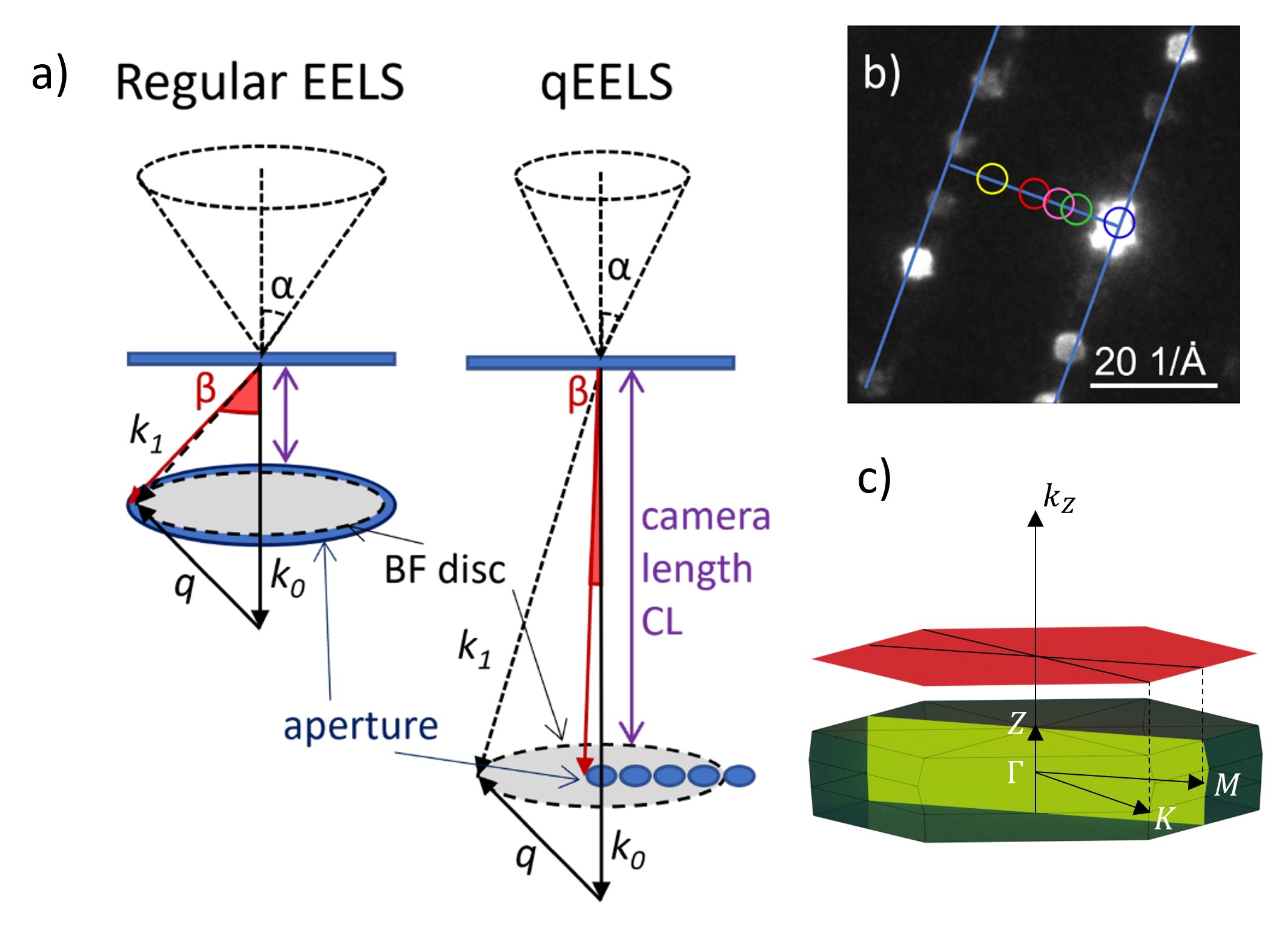}
    \caption{a) Schematic with geometry of scattering in regular EELS and momentum-resolved EELS (qEELS) where $\alpha$ is the convergence angle, $\beta$ is the collection angle, q is the momentum transfer ($q = k_1-k_0$), q($\alpha$) is the radius of the bright field disc in reciprocal space and $\Delta$q is the direction of the momentum acquired over. (b) Ronchigram of $Bi_2Se_3$ with the EELS aperture overlaid at the locations along the $\Gamma M$ direction of the first Brillouin zone centred 0, 0.7, 1.0 and 1.4 \r{A} from the centre. (c) First Brillouin zone of $Bi_2 Se_3$ with high symmetry momentum directions indicated. The shaded yellow area represents a cross-sectional sample and the red shaded are a plane view sample.}
    \label{fig:QEELSDiagram}
\end{figure}

The plasmon dispersion can be obtained from momentum-resolved EELS measurements which allow for the plasmon energy to be obtained at different momenta. This technique is an extension of standard STEM-EELS in which a focused electron beam is transmitted through a sample and collected by an EELS aperture into a spectrometer as shown in figure \ref{fig:QEELSDiagram}. In this case, instead of having a large EELS aperture to collect as many of the transmitted electrons as possible, a smaller aperture is used to further restrict the scattering angles, and hence momentum transfer range, accessed by the spectrometer \cite{Hage2013}. This is achieved with a combination of increased camera length, smaller aperture and larger convergence angle. Using a larger convergence semi-angle increases the separation of the diffraction spots allowing for the aperture to only cover a small region of the Brillouin zone in reciprocal space.
The scattering angle relates to the momentum transfer in the scattering process by

\begin{equation} \label{eq:q_angle}
    \mathbf{q}(\theta)=4\pi\sin{(\theta/2)}/\lambda,
\end{equation} 

where $\lambda$ is the electron wavelength and $\theta$ is the scattering angle of transmitted electrons. The radius of the unscattered beam without an aperture, the bright field disc, in momentum space is calculated using $\theta = \alpha$ the convergence angle. The momentum range over which the EELS spectrum is collected is found by $\theta = \beta$ the collection angle and the momentum resolution found using $\theta=\sqrt{\alpha^2+\beta^2}$, the root sums of squares of the convergence and collection angles. The resolution is determined by the relative size of the EELS aperture and diffraction pattern which results from both the collection and convergence semi-angles.
The diffraction pattern is displaced with respect to the spectrometer aperture, along specific directions in momentum space to collect data at different momentum transfer values $\Delta q$. The high symmetry $\Gamma M$ direction within the Brillouin Zone was determined using the diffraction pattern to navigate. EELS spectra were then decomposed to yield distinct contributions. A plot of the plasmon peak position versus the momentum transfer gave the plasmon dispersion relation.

 \subsection*{Density Functional Theory}
 To gain atomistic insight into the impacts of $C_{60}$ molecules on the electronic structure of the $Bi_2Se_3$, first principles calculations were performed within the framework of density functional theory (DFT), as implemented in QuantumATK \cite{Smidstrup_2020}. Linear combination of numerical atomic orbital (LCAO) basis set and generalised gradient approximation (GGA) with norm conserving pseudopotentials from PseudoDojo \cite{VANSETTEN201839} were employed. In order to obtain the Dirac cone surface states in the 2D energy-momentum relation, spin orbit coupling (SOC) through the use of fully relativistic pseudo potentials is included in the calculations. Brillouin zone integrations were performed over a grid of k points generated according to the Monkhorst Pack scheme \cite{Monkhorst1976} with a density of approximately 10 k-points per angstrom. Energy cut off of 125 Ha has been considered for discretised grid and all structural relaxation was performed with the maximum force of less than 0.02 eV \AA$^{-1}$. Van der Waals correction to the GGA functional \cite{Grimme2010} is considered to account for inter and intra molecular noncovalent of large range interaction. The slab in the supercell is infinite and periodic in the x  and y directions (parallel to the slab surface) and is finite along the z direction (normal to the slab surface). The thickness of the vacuum region along the z-direction is larger than 20 \AA to avoid any interaction between the periodic images of the neighbouring films. In the slab model calculation with asymmetric surface, an artificial macroscopic electrostatic field exists due to the periodic boundary conditions \cite{Neugebauer1992}. In order to avoid this artificial field in the $Bi_2Se_3$ with $C_{60}$ molecule on the surface, we considered Neumann and Dirichlet boundary conditions at the $C_{60}$ and $Bi_2Se_3$ sides of the slab, respectively, which provides an alternative approach for the dipole correction in the slab calculations \cite{Makov1995}.





\bmhead{Acknowledgements}

The authors would like to acknowledge the useful discussions with Prof Oscar Cespedes regarding the preparation of materials. 

\section*{Declarations}

 This work was supported by the Royal Academy of Engineering under the Research Fellowship Scheme grant number RF/201920/19/245, the Science Foundation Ireland AMBER Research Centre (SFI-12/RC/2278-P2), the European Union Horizon 2020 project ASCENT+ (grant number 871130) and EXTREME-IR (grant number 964735), the SFI/HEA Irish Centre for High-End Computing (ICHEC), the EPSRC via the NAME Programme Grant (EP/V001914/1) studentship 26045452, and the Royce Institute (EP/P022464/1 and EP/R00661X/1).






\bibliography{sn-bibliography}


\begin{thebibliography}{56}
\ifx \bisbn   \undefined \def \bisbn  #1{ISBN #1}\fi
\ifx \binits  \undefined \def \binits#1{#1}\fi
\ifx \bauthor  \undefined \def \bauthor#1{#1}\fi
\ifx \batitle  \undefined \def \batitle#1{#1}\fi
\ifx \bjtitle  \undefined \def \bjtitle#1{#1}\fi
\ifx \bvolume  \undefined \def \bvolume#1{\textbf{#1}}\fi
\ifx \byear  \undefined \def \byear#1{#1}\fi
\ifx \bissue  \undefined \def \bissue#1{#1}\fi
\ifx \bfpage  \undefined \def \bfpage#1{#1}\fi
\ifx \blpage  \undefined \def \blpage #1{#1}\fi
\ifx \burl  \undefined \def \burl#1{\textsf{#1}}\fi
\ifx \doiurl  \undefined \def \doiurl#1{\url{https://doi.org/#1}}\fi
\ifx \betal  \undefined \def \betal{\textit{et al.}}\fi
\ifx \binstitute  \undefined \def \binstitute#1{#1}\fi
\ifx \binstitutionaled  \undefined \def \binstitutionaled#1{#1}\fi
\ifx \bctitle  \undefined \def \bctitle#1{#1}\fi
\ifx \beditor  \undefined \def \beditor#1{#1}\fi
\ifx \bpublisher  \undefined \def \bpublisher#1{#1}\fi
\ifx \bbtitle  \undefined \def \bbtitle#1{#1}\fi
\ifx \bedition  \undefined \def \bedition#1{#1}\fi
\ifx \bseriesno  \undefined \def \bseriesno#1{#1}\fi
\ifx \blocation  \undefined \def \blocation#1{#1}\fi
\ifx \bsertitle  \undefined \def \bsertitle#1{#1}\fi
\ifx \bsnm \undefined \def \bsnm#1{#1}\fi
\ifx \bsuffix \undefined \def \bsuffix#1{#1}\fi
\ifx \bparticle \undefined \def \bparticle#1{#1}\fi
\ifx \barticle \undefined \def \barticle#1{#1}\fi
\bibcommenthead
\ifx \bconfdate \undefined \def \bconfdate #1{#1}\fi
\ifx \botherref \undefined \def \botherref #1{#1}\fi
\ifx \url \undefined \def \url#1{\textsf{#1}}\fi
\ifx \bchapter \undefined \def \bchapter#1{#1}\fi
\ifx \bbook \undefined \def \bbook#1{#1}\fi
\ifx \bcomment \undefined \def \bcomment#1{#1}\fi
\ifx \oauthor \undefined \def \oauthor#1{#1}\fi
\ifx \citeauthoryear \undefined \def \citeauthoryear#1{#1}\fi
\ifx \endbibitem  \undefined \def \endbibitem {}\fi
\ifx \bconflocation  \undefined \def \bconflocation#1{#1}\fi
\ifx \arxivurl  \undefined \def \arxivurl#1{\textsf{#1}}\fi
\csname PreBibitemsHook\endcsname

\bibitem[\protect\citeauthoryear{Khurgin and Sun}{2017}]{Khurgin2017}
\begin{bbook}
\bauthor{\bsnm{Khurgin}, \binits{J.B.}},
\bauthor{\bsnm{Sun}, \binits{G.}}:
In: \beditor{\bsnm{Bozhevolnyi}, \binits{S.I.}},
\beditor{\bsnm{Martin-Moreno}, \binits{L.}},
\beditor{\bsnm{Garcia-Vidal}, \binits{F.}} (eds.)
\bbtitle{Landau Damping---The Ultimate Limit of Field Confinement and Enhancement in Plasmonic Structures},
pp. \bfpage{303}--\blpage{322}.
\bpublisher{Springer},
\blocation{Cham}
(\byear{2017}).
\doiurl{10.1007/978-3-319-45820-5_13}
\end{bbook}
\endbibitem

\bibitem[\protect\citeauthoryear{Iranzo et~al.}{2018}]{Iranzo2018}
\begin{barticle}
\bauthor{\bsnm{Iranzo}, \binits{D.A.}},
\bauthor{\bsnm{Nanot}, \binits{S.}},
\bauthor{\bsnm{Dias}, \binits{E.J.C.}},
\bauthor{\bsnm{Epstein}, \binits{I.}},
\bauthor{\bsnm{Peng}, \binits{C.}},
\bauthor{\bsnm{Efetov}, \binits{D.K.}},
\bauthor{\bsnm{Lundeberg}, \binits{M.B.}},
\bauthor{\bsnm{Parret}, \binits{R.}},
\bauthor{\bsnm{Osmond}, \binits{J.}},
\bauthor{\bsnm{Hong}, \binits{J.-Y.}},
\bauthor{\bsnm{Kong}, \binits{J.}},
\bauthor{\bsnm{Englund}, \binits{D.R.}},
\bauthor{\bsnm{Peres}, \binits{N.M.R.}},
\bauthor{\bsnm{Koppens}, \binits{F.H.L.}}:
\batitle{Probing the ultimate plasmon confinement limits with a van der waals heterostructure}.
\bjtitle{Science}
\bvolume{360}(\bissue{6386}),
\bfpage{291}--\blpage{295}
(\byear{2018})
\doiurl{10.1126/science.aar8438}
\end{barticle}
\endbibitem

\bibitem[\protect\citeauthoryear{Li et~al.}{2017}]{Li2017}
\begin{barticle}
\bauthor{\bsnm{Li}, \binits{Y.}},
\bauthor{\bsnm{Li}, \binits{Z.}},
\bauthor{\bsnm{Chi}, \binits{C.}},
\bauthor{\bsnm{Shan}, \binits{H.}},
\bauthor{\bsnm{Zheng}, \binits{L.}},
\bauthor{\bsnm{Fang}, \binits{Z.}}:
\batitle{Plasmonics of 2d nanomaterials: Properties and applications}.
\bjtitle{Advanced Science}
\bvolume{4}(\bissue{8}),
\bfpage{1600430}
(\byear{2017})
\doiurl{10.1002/advs.201600430}
{\href{https://arxiv.org/abs/https://onlinelibrary.wiley.com/doi/pdf/10.1002/advs.201600430}{{https://onlinelibrary.wiley.com/doi/pdf/10.1002/advs.201600430}}}
\end{barticle}
\endbibitem

\bibitem[\protect\citeauthoryear{Grigorenko et~al.}{2012}]{Grigorenko2012}
\begin{barticle}
\bauthor{\bsnm{Grigorenko}, \binits{A.N.}},
\bauthor{\bsnm{Polini}, \binits{M.}},
\bauthor{\bsnm{Novoselov}, \binits{K.S.}}:
\batitle{Graphene plasmonics}.
\bjtitle{Nature Photonics}
\bvolume{6},
\bfpage{749}--\blpage{758}
(\byear{2012})
\doiurl{10.1038/nphoton.2012.262}
\end{barticle}
\endbibitem

\bibitem[\protect\citeauthoryear{Lai et~al.}{2014}]{Lai2014}
\begin{barticle}
\bauthor{\bsnm{Lai}, \binits{Y.-P.}},
\bauthor{\bsnm{Lin}, \binits{I.-T.}},
\bauthor{\bsnm{Wu}, \binits{K.-H.}},
\bauthor{\bsnm{Liu}, \binits{J.-M.}}:
\batitle{Plasmonics in topological insulators}.
\bjtitle{Nanomaterials and Nanotechnology}
\bvolume{4},
\bfpage{13}
(\byear{2014})
\doiurl{10.5772/58558}
{\href{https://arxiv.org/abs/https://doi.org/10.5772/58558}{{https://doi.org/10.5772/58558}}}
\end{barticle}
\endbibitem

\bibitem[\protect\citeauthoryear{Stauber}{2014}]{Stauber2014}
\begin{barticle}
\bauthor{\bsnm{Stauber}, \binits{T.}}:
\batitle{Plasmonics in dirac systems: from graphene to topological insulators}.
\bjtitle{Journal of Physics: Condensed Matter}
\bvolume{26}(\bissue{12}),
\bfpage{123201}
(\byear{2014})
\doiurl{10.1088/0953-8984/26/12/123201}
\end{barticle}
\endbibitem

\bibitem[\protect\citeauthoryear{yin et~al.}{2017}]{Yin2017}
\begin{botherref}
\oauthor{\bsnm{yin}, \binits{J.}},
\oauthor{\bsnm{Krishnamoorthy}, \binits{H.N.}},
\oauthor{\bsnm{Adamo}, \binits{G.}},
\oauthor{\bsnm{Dubrovkin}, \binits{A.}},
\oauthor{\bsnm{Chong}, \binits{Y.}},
\oauthor{\bsnm{Zheludev}, \binits{N.I.}},
\oauthor{\bsnm{Soci}, \binits{C.}}:
Plasmonics of topological insulators at optical frequencies.
NPG Asia Materials
\textbf{9}(e425)
(2017)
\doiurl{10.1038/am.2017.149}
\end{botherref}
\endbibitem

\bibitem[\protect\citeauthoryear{Di~Pietro et~al.}{2020}]{DiPietro202}
\begin{barticle}
\bauthor{\bsnm{Di~Pietro}, \binits{P.}},
\bauthor{\bsnm{Adhlakha}, \binits{N.}},
\bauthor{\bsnm{Piccirilli}, \binits{F.}},
\bauthor{\bsnm{Di~Gaspare}, \binits{A.}},
\bauthor{\bsnm{Moon}, \binits{J.}},
\bauthor{\bsnm{Oh}, \binits{S.}},
\bauthor{\bsnm{Di~Mitri}, \binits{S.}},
\bauthor{\bsnm{Spampinati}, \binits{S.}},
\bauthor{\bsnm{Perucchi}, \binits{A.}},
\bauthor{\bsnm{Lupi}, \binits{S.}}:
\batitle{Terahertz tuning of dirac plasmons in ${\mathrm{bi}}_{2}{\mathrm{se}}_{3}$ topological insulator}.
\bjtitle{Phys. Rev. Lett.}
\bvolume{124},
\bfpage{226403}
(\byear{2020})
\doiurl{10.1103/PhysRevLett.124.226403}
\end{barticle}
\endbibitem

\bibitem[\protect\citeauthoryear{Hasan and Kane}{2010}]{Hasan2010}
\begin{barticle}
\bauthor{\bsnm{Hasan}, \binits{M.Z.}},
\bauthor{\bsnm{Kane}, \binits{C.L.}}:
\batitle{Colloquium: Topological insulators}.
\bjtitle{Rev. Mod. Phys.}
\bvolume{82},
\bfpage{3045}--\blpage{3067}
(\byear{2010})
\doiurl{10.1103/RevModPhys.82.3045}
\end{barticle}
\endbibitem

\bibitem[\protect\citeauthoryear{Pietro et~al.}{2013}]{DiPietro2013}
\begin{barticle}
\bauthor{\bsnm{Pietro}, \binits{P.D.}},
\bauthor{\bsnm{Ortolani}, \binits{M.}},
\bauthor{\bsnm{Limaj}, \binits{O.}},
\bauthor{\bsnm{Gaspare}, \binits{A.D.}},
\bauthor{\bsnm{Giliberti}, \binits{V.}},
\bauthor{\bsnm{Giorgianni}, \binits{F.}},
\bauthor{\bsnm{Brahlek}, \binits{M.}},
\bauthor{\bsnm{Bansal}, \binits{N.}},
\bauthor{\bsnm{Koirala}, \binits{N.}},
\bauthor{\bsnm{Oh}, \binits{S.}},
\bauthor{\bsnm{Calvani}, \binits{P.}},
\bauthor{\bsnm{Lupi}, \binits{S.}}:
\batitle{Observation of dirac plasmons in a topological insulator}.
\bjtitle{Nature Nanotech.}
\bvolume{8},
\bfpage{556}--\blpage{560}
(\byear{2013})
\doiurl{10.1038/nnano.2013.134}
\end{barticle}
\endbibitem

\bibitem[\protect\citeauthoryear{Stauber et~al.}{2013}]{Stauber2013}
\begin{barticle}
\bauthor{\bsnm{Stauber}, \binits{T.}},
\bauthor{\bsnm{G\'omez-Santos}, \binits{G.}},
\bauthor{\bsnm{Brey}, \binits{L.}}:
\batitle{Spin-charge separation of plasmonic excitations in thin topological insulators}.
\bjtitle{Phys. Rev. B}
\bvolume{88},
\bfpage{205427}
(\byear{2013})
\doiurl{10.1103/PhysRevB.88.205427}
\end{barticle}
\endbibitem

\bibitem[\protect\citeauthoryear{Raghu et~al.}{2010}]{Raghu2010}
\begin{barticle}
\bauthor{\bsnm{Raghu}, \binits{S.}},
\bauthor{\bsnm{Chung}, \binits{S.B.}},
\bauthor{\bsnm{Qi}, \binits{X.-L.}},
\bauthor{\bsnm{Zhang}, \binits{S.-C.}}:
\batitle{Collective modes of a helical liquid}.
\bjtitle{Phys. Rev. Lett.}
\bvolume{104},
\bfpage{116401}
(\byear{2010})
\doiurl{10.1103/PhysRevLett.104.116401}
\end{barticle}
\endbibitem

\bibitem[\protect\citeauthoryear{Bianchi et~al.}{2010}]{Bianchi2010}
\begin{botherref}
\oauthor{\bsnm{Bianchi}, \binits{M.}},
\oauthor{\bsnm{Guan}, \binits{D.}},
\oauthor{\bsnm{Bao}, \binits{S.}},
\oauthor{\bsnm{Mi}, \binits{J.}},
\oauthor{\bsnm{Iversen}, \binits{B.B.}},
\oauthor{\bsnm{King}, \binits{P.D.C.}},
\oauthor{\bsnm{Hoffmann}, \binits{P.}}:
Coexistence of the topological state and a two-dimensional electron gas on the surface of bi2se3.
Nature Comm.
\textbf{1}
(2010)
\doiurl{10.1038/ncomms1131}
\end{botherref}
\endbibitem

\bibitem[\protect\citeauthoryear{Bianchi et~al.}{2012}]{Bianchi2012}
\begin{barticle}
\bauthor{\bsnm{Bianchi}, \binits{M.}},
\bauthor{\bsnm{Hatch}, \binits{R.C.}},
\bauthor{\bsnm{Li}, \binits{Z.}},
\bauthor{\bsnm{Hofmann}, \binits{P.}},
\bauthor{\bsnm{Song}, \binits{F.}},
\bauthor{\bsnm{Mi}, \binits{J.}},
\bauthor{\bsnm{Iversen}, \binits{B.B.}},
\bauthor{\bsnm{Abd~El-Fattah}, \binits{Z.M.}},
\bauthor{\bsnm{Löptien}, \binits{P.}},
\bauthor{\bsnm{Zhou}, \binits{L.}},
\bauthor{\bsnm{Khajetoorians}, \binits{A.A.}},
\bauthor{\bsnm{Wiebe}, \binits{J.}},
\bauthor{\bsnm{Wiesendanger}, \binits{R.}},
\bauthor{\bsnm{Wells}, \binits{J.W.}}:
\batitle{Robust surface doping of bi2se3 by rubidium intercalation}.
\bjtitle{ACS Nano}
\bvolume{6}(\bissue{8}),
\bfpage{7009}--\blpage{7015}
(\byear{2012})
\doiurl{10.1021/nn3021822} .
\bcomment{PMID: 22838508}
\end{barticle}
\endbibitem

\bibitem[\protect\citeauthoryear{Kitazawa et~al.}{2020}]{Kitazawa2020}
\begin{barticle}
\bauthor{\bsnm{Kitazawa}, \binits{T.}},
\bauthor{\bsnm{Yaji}, \binits{K.}},
\bauthor{\bsnm{Shimozawa}, \binits{K.}},
\bauthor{\bsnm{Kondo}, \binits{H.}},
\bauthor{\bsnm{Yamanaka}, \binits{T.}},
\bauthor{\bsnm{Yaguchi}, \binits{H.}},
\bauthor{\bsnm{Ishida}, \binits{Y.}},
\bauthor{\bsnm{Kuroda}, \binits{K.}},
\bauthor{\bsnm{Harasawa}, \binits{A.}},
\bauthor{\bsnm{Iwahashi}, \binits{T.}},
\bauthor{\bsnm{Ouchi}, \binits{Y.}},
\bauthor{\bsnm{Komori}, \binits{F.}},
\bauthor{\bsnm{Shin}, \binits{S.}},
\bauthor{\bsnm{Kanai}, \binits{K.}}:
\batitle{Topological surface state of bi2se3 modified by adsorption of organic donor molecule tetrathianaphthacene}.
\bjtitle{Advanced Materials Interfaces}
\bvolume{7}(\bissue{14}),
\bfpage{2000524}
(\byear{2020})
\doiurl{10.1002/admi.202000524}
\end{barticle}
\endbibitem

\bibitem[\protect\citeauthoryear{Wang et~al.}{2014}]{Wang2014}
\begin{barticle}
\bauthor{\bsnm{Wang}, \binits{J.}},
\bauthor{\bsnm{Hewitt}, \binits{A.S.}},
\bauthor{\bsnm{Kumar}, \binits{R.}},
\bauthor{\bsnm{Boltersdorf}, \binits{J.}},
\bauthor{\bsnm{Guan}, \binits{T.}},
\bauthor{\bsnm{Hunte}, \binits{F.}},
\bauthor{\bsnm{Maggard}, \binits{P.A.}},
\bauthor{\bsnm{Brom}, \binits{J.E.}},
\bauthor{\bsnm{Redwing}, \binits{J.M.}},
\bauthor{\bsnm{Dougherty}, \binits{D.B.}}:
\batitle{Molecular doping control at a topological insulator surface: F4-tcnq on bi2se3}.
\bjtitle{The Journal of Physical Chemistry C}
\bvolume{118}(\bissue{27}),
\bfpage{14860}--\blpage{14865}
(\byear{2014})
\doiurl{10.1021/jp412690h}
\end{barticle}
\endbibitem

\bibitem[\protect\citeauthoryear{Caputo et~al.}{2016}]{Caputo2016}
\begin{barticle}
\bauthor{\bsnm{Caputo}, \binits{M.}},
\bauthor{\bsnm{Panighel}, \binits{M.}},
\bauthor{\bsnm{Lisi}, \binits{S.}},
\bauthor{\bsnm{Khalil}, \binits{L.}},
\bauthor{\bsnm{Santo}, \binits{G.D.}},
\bauthor{\bsnm{Papalazarou}, \binits{E.}},
\bauthor{\bsnm{Hruban}, \binits{A.}},
\bauthor{\bsnm{Konczykowski}, \binits{M.}},
\bauthor{\bsnm{Krusin-Elbaum}, \binits{L.}},
\bauthor{\bsnm{Aliev}, \binits{Z.S.}},
\bauthor{\bsnm{Babanly}, \binits{M.B.}},
\bauthor{\bsnm{Otrokov}, \binits{M.M.}},
\bauthor{\bsnm{Politano}, \binits{A.}},
\bauthor{\bsnm{Chulkov}, \binits{E.V.}},
\bauthor{\bsnm{Arnau}, \binits{A.}},
\bauthor{\bsnm{Marinova}, \binits{V.}},
\bauthor{\bsnm{Das}, \binits{P.K.}},
\bauthor{\bsnm{Fujii}, \binits{J.}},
\bauthor{\bsnm{Vobornik}, \binits{I.}},
\bauthor{\bsnm{Perfetti}, \binits{L.}},
\bauthor{\bsnm{Mugarza}, \binits{A.}},
\bauthor{\bsnm{Goldoni}, \binits{A.}},
\bauthor{\bsnm{Marsi}, \binits{M.}}:
\batitle{Manipulating the topological interface by molecular adsorbates: Adsorption of co-phthalocyanine on bi2se3}.
\bjtitle{Nano Letters}
\bvolume{16}(\bissue{6}),
\bfpage{3409}--\blpage{3414}
(\byear{2016})
\doiurl{10.1021/acs.nanolett.5b02635} .
\bcomment{PMID: 27010705}
\end{barticle}
\endbibitem

\bibitem[\protect\citeauthoryear{Wei et~al.}{2022}]{Wei2022}
\begin{barticle}
\bauthor{\bsnm{Wei}, \binits{T.}},
\bauthor{\bsnm{Liu}, \binits{Y.}},
\bauthor{\bsnm{Cui}, \binits{P.}},
\bauthor{\bsnm{Li}, \binits{X.}},
\bauthor{\bsnm{Zhang}, \binits{Z.}}:
\batitle{Emergent optical plasmons at the surface of a doped three-dimensional topological insulator}.
\bjtitle{Phys. Rev. B}
\bvolume{105},
\bfpage{205408}
(\byear{2022})
\doiurl{10.1103/PhysRevB.105.205408}
\end{barticle}
\endbibitem

\bibitem[\protect\citeauthoryear{Whitcher et~al.}{2020}]{whitcher2020correlated}
\begin{barticle}
\bauthor{\bsnm{Whitcher}, \binits{T.J.}},
\bauthor{\bsnm{Silly}, \binits{M.G.}},
\bauthor{\bsnm{Yang}, \binits{M.}},
\bauthor{\bsnm{Das}, \binits{P.K.}},
\bauthor{\bsnm{Peyrot}, \binits{D.}},
\bauthor{\bsnm{Chi}, \binits{X.}},
\bauthor{\bsnm{Eddrief}, \binits{M.}},
\bauthor{\bsnm{Moon}, \binits{J.}},
\bauthor{\bsnm{Oh}, \binits{S.}},
\bauthor{\bsnm{Castro-Neto}, \binits{A.H.}}, \betal:
\batitle{Correlated plasmons in the topological insulator bi2se3 induced by long-range electron correlations}.
\bjtitle{NPG Asia Materials}
\bvolume{12}(\bissue{1}),
\bfpage{37}
(\byear{2020})
\end{barticle}
\endbibitem

\bibitem[\protect\citeauthoryear{Jakobs et~al.}{2015}]{Jakobs2015}
\begin{barticle}
\bauthor{\bsnm{Jakobs}, \binits{S.}},
\bauthor{\bsnm{Narayan}, \binits{A.}},
\bauthor{\bsnm{Stadtmüller}, \binits{B.}},
\bauthor{\bsnm{Droghetti}, \binits{A.}},
\bauthor{\bsnm{Rungger}, \binits{I.}},
\bauthor{\bsnm{Hor}, \binits{Y.S.}},
\bauthor{\bsnm{Klyatskaya}, \binits{S.}},
\bauthor{\bsnm{Jungkenn}, \binits{D.}},
\bauthor{\bsnm{Stöckl}, \binits{J.}},
\bauthor{\bsnm{Laux}, \binits{M.}},
\bauthor{\bsnm{Monti}, \binits{O.L.A.}},
\bauthor{\bsnm{Aeschlimann}, \binits{M.}},
\bauthor{\bsnm{Cava}, \binits{R.J.}},
\bauthor{\bsnm{Ruben}, \binits{M.}},
\bauthor{\bsnm{Mathias}, \binits{S.}},
\bauthor{\bsnm{Sanvito}, \binits{S.}},
\bauthor{\bsnm{Cinchetti}, \binits{M.}}:
\batitle{Controlling the spin texture of topological insulators by rational design of organic molecules}.
\bjtitle{Nano Letters}
\bvolume{15}(\bissue{9}),
\bfpage{6022}--\blpage{6029}
(\byear{2015})
\doiurl{10.1021/acs.nanolett.5b02213} .
\bcomment{PMID: 26262825}
\end{barticle}
\endbibitem

\bibitem[\protect\citeauthoryear{Cai et~al.}{2018}]{cai2018independence}
\begin{barticle}
\bauthor{\bsnm{Cai}, \binits{S.}},
\bauthor{\bsnm{Guo}, \binits{J.}},
\bauthor{\bsnm{Sidorov}, \binits{V.A.}},
\bauthor{\bsnm{Zhou}, \binits{Y.}},
\bauthor{\bsnm{Wang}, \binits{H.}},
\bauthor{\bsnm{Lin}, \binits{G.}},
\bauthor{\bsnm{Li}, \binits{X.}},
\bauthor{\bsnm{Li}, \binits{Y.}},
\bauthor{\bsnm{Yang}, \binits{K.}},
\bauthor{\bsnm{Li}, \binits{A.}}, \betal:
\batitle{Independence of topological surface state and bulk conductance in three-dimensional topological insulators}.
\bjtitle{Npj Quantum Materials}
\bvolume{3}(\bissue{1}),
\bfpage{62}
(\byear{2018})
\end{barticle}
\endbibitem

\bibitem[\protect\citeauthoryear{Jia et~al.}{2017}]{Jia2017}
\begin{botherref}
\oauthor{\bsnm{Jia}, \binits{X.}},
\oauthor{\bsnm{Zhang}, \binits{S.}},
\oauthor{\bsnm{Sankar}, \binits{R.}},
\oauthor{\bsnm{Chou}, \binits{F.C.}},
\oauthor{\bsnm{Wang}, \binits{W.}},
\oauthor{\bsnm{Kempa}, \binits{K.}},
\oauthor{\bsnm{Plummer}, \binits{E.W.}},
\oauthor{\bsnm{Zhang}, \binits{J.}},
\oauthor{\bsnm{Zhu}, \binits{X.}},
\oauthor{\bsnm{Guo}, \binits{J.}}:
Anomalous acoustic plasmon mode from topologically protected states.
Physical Review Letters
\textbf{119}
(2017)
\doiurl{10.1103/PhysRevLett.119.136805}
\end{botherref}
\endbibitem

\bibitem[\protect\citeauthoryear{Liou et~al.}{2013}]{Liou2013}
\begin{botherref}
\oauthor{\bsnm{Liou}, \binits{S.C.}},
\oauthor{\bsnm{Chu}, \binits{M.W.}},
\oauthor{\bsnm{Sankar}, \binits{R.}},
\oauthor{\bsnm{Huang}, \binits{F.T.}},
\oauthor{\bsnm{Shu}, \binits{G.J.}},
\oauthor{\bsnm{Chou}, \binits{F.C.}},
\oauthor{\bsnm{Chen}, \binits{C.H.}}:
Plasmons dispersion and nonvertical interband transitions in single crystal bi2se3 investigated by electron energy-loss spectroscopy.
Physical Review B - Condensed Matter and Materials Physics
\textbf{87}
(2013)
\doiurl{10.1103/PhysRevB.87.085126}
\end{botherref}
\endbibitem

\bibitem[\protect\citeauthoryear{Politano and Chiarello}{2014}]{politano2014plasmon}
\begin{barticle}
\bauthor{\bsnm{Politano}, \binits{A.}},
\bauthor{\bsnm{Chiarello}, \binits{G.}}:
\batitle{Plasmon modes in graphene: status and prospect}.
\bjtitle{Nanoscale}
\bvolume{6}(\bissue{19}),
\bfpage{10927}--\blpage{10940}
(\byear{2014})
\end{barticle}
\endbibitem

\bibitem[\protect\citeauthoryear{Despoja et~al.}{2013}]{despoja2013two}
\begin{barticle}
\bauthor{\bsnm{Despoja}, \binits{V.}},
\bauthor{\bsnm{Novko}, \binits{D.}},
\bauthor{\bsnm{Dekani{\'c}}, \binits{K.}},
\bauthor{\bsnm{{\v{S}}unji{\'c}}, \binits{M.}},
\bauthor{\bsnm{Maru{\v{s}}i{\'c}}, \binits{L.}}:
\batitle{Two-dimensional and $\pi$ plasmon spectra in pristine and doped graphene}.
\bjtitle{Physical Review B—Condensed Matter and Materials Physics}
\bvolume{87}(\bissue{7}),
\bfpage{075447}
(\byear{2013})
\end{barticle}
\endbibitem

\bibitem[\protect\citeauthoryear{Latzke et~al.}{2019}]{Latzke2019}
\begin{barticle}
\bauthor{\bsnm{Latzke}, \binits{D.W.}},
\bauthor{\bsnm{Ojeda-Aristizabal}, \binits{C.}},
\bauthor{\bsnm{Griffin}, \binits{S.M.}},
\bauthor{\bsnm{Denlinger}, \binits{J.D.}},
\bauthor{\bsnm{Neaton}, \binits{J.B.}},
\bauthor{\bsnm{Zettl}, \binits{A.}},
\bauthor{\bsnm{Lanzara}, \binits{A.}}:
\batitle{Observation of highly dispersive bands in pure thin film ${\mathrm{c}}_{60}$}.
\bjtitle{Phys. Rev. B}
\bvolume{99},
\bfpage{045425}
(\byear{2019})
\doiurl{10.1103/PhysRevB.99.045425}
\end{barticle}
\endbibitem

\bibitem[\protect\citeauthoryear{Pistore et~al.}{2024}]{pistore2024terahertz}
\begin{barticle}
\bauthor{\bsnm{Pistore}, \binits{V.}},
\bauthor{\bsnm{Viti}, \binits{L.}},
\bauthor{\bsnm{Schiattarella}, \binits{C.}},
\bauthor{\bsnm{Riccardi}, \binits{E.}},
\bauthor{\bsnm{Knox}, \binits{C.S.}},
\bauthor{\bsnm{Yagmur}, \binits{A.}},
\bauthor{\bsnm{Burton}, \binits{J.J.}},
\bauthor{\bsnm{Sasaki}, \binits{S.}},
\bauthor{\bsnm{Davies}, \binits{A.G.}},
\bauthor{\bsnm{Linfield}, \binits{E.H.}}, \betal:
\batitle{Terahertz plasmon polaritons in large area bi2se3 topological insulators}.
\bjtitle{Advanced Optical Materials}
\bvolume{12}(\bissue{4}),
\bfpage{2301673}
(\byear{2024})
\end{barticle}
\endbibitem

\bibitem[\protect\citeauthoryear{Ostling et~al.}{1993}]{Ostling1993}
\begin{barticle}
\bauthor{\bsnm{Ostling}, \binits{D.}},
\bauthor{\bsnm{Apell}, \binits{P.}},
\bauthor{\bsnm{Rosén}, \binits{A.}}:
\batitle{Surface plasmons of c60}.
\bjtitle{Zeitschrift fur Physik D Atoms, Molecules and Clusters}
\bvolume{26},
\bfpage{282}--\blpage{284}
(\byear{1993})
\doiurl{10.1007/BF01425691}
\end{barticle}
\endbibitem

\bibitem[\protect\citeauthoryear{Barton and Eberlein}{1991}]{Barton1991}
\begin{barticle}
\bauthor{\bsnm{Barton}, \binits{G.}},
\bauthor{\bsnm{Eberlein}, \binits{C.}}:
\batitle{{Plasma spectroscopy proposed for C60 and C70}}.
\bjtitle{The Journal of Chemical Physics}
\bvolume{95}(\bissue{3}),
\bfpage{1512}--\blpage{1517}
(\byear{1991})
\doiurl{10.1063/1.461065}
\end{barticle}
\endbibitem

\bibitem[\protect\citeauthoryear{Bolognesi et~al.}{2012}]{BOLOGNESI2012}
\begin{botherref}
\oauthor{\bsnm{Bolognesi}, \binits{P.}},
\oauthor{\bsnm{Avaldi}, \binits{L.}},
\oauthor{\bsnm{Rucco}, \binits{A.}},
\oauthor{\bsnm{Verkhovstev}, \binits{A.}},
\oauthor{\bsnm{Korol}, \binits{A.V.}},
\oauthor{\bsnm{Solov'yov}, \binits{A.V.}}:
Collective excitations in the electron energy loss spectra of c60.
Eur. Phys. J. D
\textbf{66}(254)
(2012)
\doiurl{10.1140/epjd/e2012-30178-1}
\end{botherref}
\endbibitem

\bibitem[\protect\citeauthoryear{Hansen et~al.}{1991}]{HANSEN1991}
\begin{barticle}
\bauthor{\bsnm{Hansen}, \binits{P.L.}},
\bauthor{\bsnm{Fallon}, \binits{P.J.}},
\bauthor{\bsnm{Krätschmer}, \binits{W.}}:
\batitle{An eels study of fullerite — c60/c70}.
\bjtitle{Chemical Physics Letters}
\bvolume{181}(\bissue{4}),
\bfpage{367}--\blpage{372}
(\byear{1991})
\doiurl{10.1016/0009-2614(91)80086-D}
\end{barticle}
\endbibitem

\bibitem[\protect\citeauthoryear{Gorokhov et~al.}{1996}]{GOROKHOV1996}
\begin{barticle}
\bauthor{\bsnm{Gorokhov}, \binits{D.A.}},
\bauthor{\bsnm{Suris}, \binits{R.A.}},
\bauthor{\bsnm{Cheianov}, \binits{V.V.}}:
\batitle{Electron-energy-loss spectroscopy of the c60 molecule}.
\bjtitle{Physics Letters A}
\bvolume{223}(\bissue{1}),
\bfpage{116}--\blpage{122}
(\byear{1996})
\doiurl{10.1016/S0375-9601(96)00707-4}
\end{barticle}
\endbibitem

\bibitem[\protect\citeauthoryear{Gillet and Ealet}{1992}]{GILLET1992}
\begin{barticle}
\bauthor{\bsnm{Gillet}, \binits{E.}},
\bauthor{\bsnm{Ealet}, \binits{B.}}:
\batitle{Characterization of sapphire surfaces by electron energy-loss spectroscopy}.
\bjtitle{Surface Science}
\bvolume{273}(\bissue{3}),
\bfpage{427}--\blpage{436}
(\byear{1992})
\doiurl{10.1016/0039-6028(92)90079-L}
\end{barticle}
\endbibitem

\bibitem[\protect\citeauthoryear{Ealet and Gillet}{1993}]{GILLET1993}
\begin{barticle}
\bauthor{\bsnm{Ealet}, \binits{B.}},
\bauthor{\bsnm{Gillet}, \binits{E.}}:
\batitle{Palladium alumina interface: influence of the oxide stoichiometry studied by eels and xps}.
\bjtitle{Surface Science}
\bvolume{281}(\bissue{1}),
\bfpage{91}--\blpage{101}
(\byear{1993})
\doiurl{10.1016/0039-6028(93)90858-H}
\end{barticle}
\endbibitem

\bibitem[\protect\citeauthoryear{Maclean et~al.}{2001}]{Maclean2001}
\begin{botherref}
\oauthor{\bsnm{Maclean}, \binits{E.D.W.}},
\oauthor{\bsnm{Craven}, \binits{A.J.}},
\oauthor{\bsnm{W}, \binits{M.D.}}:
Valence losses at interfaces in aluminium alloys.
Inst. Phys. Conf. Ser
\textbf{168}
(2001)
\end{botherref}
\endbibitem

\bibitem[\protect\citeauthoryear{Konečná et~al.}{2018}]{Konecna2018}
\begin{botherref}
\oauthor{\bsnm{Konečná}, \binits{A.}},
\oauthor{\bsnm{Venkatraman}, \binits{K.}},
\oauthor{\bsnm{March}, \binits{K.}},
\oauthor{\bsnm{Crozier}, \binits{P.A.}},
\oauthor{\bsnm{Hillenbrand}, \binits{R.}},
\oauthor{\bsnm{Rez}, \binits{P.}},
\oauthor{\bsnm{Aizpurua}, \binits{J.}}:
Vibrational electron energy loss spectroscopy in truncated dielectric slabs.
Physical Review B
\textbf{98}
(2018)
\doiurl{10.1103/PhysRevB.98.205409}
\end{botherref}
\endbibitem

\bibitem[\protect\citeauthoryear{Krivanek et~al.}{2019}]{KRIVANEK2019}
\begin{barticle}
\bauthor{\bsnm{Krivanek}, \binits{O.L.}},
\bauthor{\bsnm{Dellby}, \binits{N.}},
\bauthor{\bsnm{Hachtel}, \binits{J.A.}},
\bauthor{\bsnm{Idrobo}, \binits{J.-C.}},
\bauthor{\bsnm{Hotz}, \binits{M.T.}},
\bauthor{\bsnm{Plotkin-Swing}, \binits{B.}},
\bauthor{\bsnm{Bacon}, \binits{N.J.}},
\bauthor{\bsnm{Bleloch}, \binits{A.L.}},
\bauthor{\bsnm{Corbin}, \binits{G.J.}},
\bauthor{\bsnm{Hoffman}, \binits{M.V.}},
\bauthor{\bsnm{Meyer}, \binits{C.E.}},
\bauthor{\bsnm{Lovejoy}, \binits{T.C.}}:
\batitle{Progress in ultrahigh energy resolution eels}.
\bjtitle{Ultramicroscopy}
\bvolume{203},
\bfpage{60}--\blpage{67}
(\byear{2019})
\doiurl{10.1016/j.ultramic.2018.12.006} .
\bcomment{75th Birthday of Christian Colliex, 85th Birthday of Archie Howie, and 75th Birthday of Hannes Lichte / PICO 2019 - Fifth Conference on Frontiers of Aberration Corrected Electron Microscopy}
\end{barticle}
\endbibitem

\bibitem[\protect\citeauthoryear{Konečná et~al.}{2018}]{Konecna2018_SEELS}
\begin{barticle}
\bauthor{\bsnm{Konečná}, \binits{A.}},
\bauthor{\bsnm{Neuman}, \binits{T.}},
\bauthor{\bsnm{Aizpurua}, \binits{J.}},
\bauthor{\bsnm{Hillenbrand}, \binits{R.}}:
\batitle{Surface-enhanced molecular electron energy loss spectroscopy}.
\bjtitle{ACS Nano}
\bvolume{12}(\bissue{5}),
\bfpage{4775}--\blpage{4786}
(\byear{2018})
\doiurl{10.1021/acsnano.8b01481} .
\bcomment{PMID: 29641179}
\end{barticle}
\endbibitem

\bibitem[\protect\citeauthoryear{Shin et~al.}{2011}]{shin2011control}
\begin{botherref}
\oauthor{\bsnm{Shin}, \binits{S.}},
\oauthor{\bsnm{Kim}, \binits{N.}},
\oauthor{\bsnm{Kim}, \binits{J.}},
\oauthor{\bsnm{Kim}, \binits{K.}},
\oauthor{\bsnm{Noh}, \binits{D.}},
\oauthor{\bsnm{Kim}, \binits{K.S.}},
\oauthor{\bsnm{Chung}, \binits{J.}}:
Control of the $\pi$ plasmon in a single layer graphene by charge doping.
Applied Physics Letters
\textbf{99}(8)
(2011)
\end{botherref}
\endbibitem

\bibitem[\protect\citeauthoryear{Nagashima et~al.}{1992}]{NAGASHIMA1992}
\begin{barticle}
\bauthor{\bsnm{Nagashima}, \binits{A.}},
\bauthor{\bsnm{Nuka}, \binits{K.}},
\bauthor{\bsnm{Itoh}, \binits{H.}},
\bauthor{\bsnm{Ichinokawa}, \binits{T.}},
\bauthor{\bsnm{Oshima}, \binits{C.}},
\bauthor{\bsnm{Otani}, \binits{S.}},
\bauthor{\bsnm{Ishizawa}, \binits{Y.}}:
\batitle{Two-dimensional plasmons in monolayer graphite}.
\bjtitle{Solid State Communications}
\bvolume{83}(\bissue{8}),
\bfpage{581}--\blpage{585}
(\byear{1992})
\doiurl{10.1016/0038-1098(92)90656-T}
\end{barticle}
\endbibitem

\bibitem[\protect\citeauthoryear{Liou et~al.}{2015}]{Liou2015}
\begin{botherref}
\oauthor{\bsnm{Liou}, \binits{S.C.}},
\oauthor{\bsnm{Shie}, \binits{C.S.}},
\oauthor{\bsnm{Chen}, \binits{C.H.}},
\oauthor{\bsnm{Breitwieser}, \binits{R.}},
\oauthor{\bsnm{Pai}, \binits{W.W.}},
\oauthor{\bsnm{Guo}, \binits{G.Y.}},
\oauthor{\bsnm{Chu}, \binits{M.W.}}:
$\pi$- plasmon dispersion in free-standing graphene by momentum resolved electron energy-loss spectroscopy.
Physical Review B - Condensed Matter and Materials Physics
\textbf{91}
(2015)
\doiurl{10.1103/PhysRevB.91.045418}
\end{botherref}
\endbibitem

\bibitem[\protect\citeauthoryear{Stern}{1967}]{Stern1967}
\begin{barticle}
\bauthor{\bsnm{Stern}, \binits{F.}}:
\batitle{Polarizability of a two-dimensional electron gas}.
\bjtitle{Phys. Rev. Lett.}
\bvolume{18},
\bfpage{546}--\blpage{548}
(\byear{1967})
\doiurl{10.1103/PhysRevLett.18.546}
\end{barticle}
\endbibitem

\bibitem[\protect\citeauthoryear{Kataura et~al.}{1997}]{kataura1997dielectric}
\begin{barticle}
\bauthor{\bsnm{Kataura}, \binits{H.}},
\bauthor{\bsnm{Endo}, \binits{Y.}},
\bauthor{\bsnm{Achiba}, \binits{Y.}},
\bauthor{\bsnm{Kikuchi}, \binits{K.}},
\bauthor{\bsnm{Hanyu}, \binits{T.}},
\bauthor{\bsnm{Yamaguchi}, \binits{S.}}:
\batitle{Dielectric constants of c60 and c70 thin films}.
\bjtitle{Journal of Physics and Chemistry of Solids}
\bvolume{58}(\bissue{11}),
\bfpage{1913}--\blpage{1917}
(\byear{1997})
\end{barticle}
\endbibitem

\bibitem[\protect\citeauthoryear{Tang and Fu}{2014}]{tang2014strain}
\begin{barticle}
\bauthor{\bsnm{Tang}, \binits{E.}},
\bauthor{\bsnm{Fu}, \binits{L.}}:
\batitle{Strain-induced partially flat band, helical snake states and interface superconductivity in topological crystalline insulators}.
\bjtitle{Nature Physics}
\bvolume{10}(\bissue{12}),
\bfpage{964}--\blpage{969}
(\byear{2014})
\end{barticle}
\endbibitem

\bibitem[\protect\citeauthoryear{Cinchetti et~al.}{2017}]{cinchetti2017activating}
\begin{barticle}
\bauthor{\bsnm{Cinchetti}, \binits{M.}},
\bauthor{\bsnm{Dediu}, \binits{V.A.}},
\bauthor{\bsnm{Hueso}, \binits{L.E.}}:
\batitle{Activating the molecular spinterface}.
\bjtitle{Nature materials}
\bvolume{16}(\bissue{5}),
\bfpage{507}--\blpage{515}
(\byear{2017})
\end{barticle}
\endbibitem

\bibitem[\protect\citeauthoryear{Knox et~al.}{2022}]{Knox2022}
\begin{barticle}
\bauthor{\bsnm{Knox}, \binits{C.S.}},
\bauthor{\bsnm{Vaughan}, \binits{M.T.}},
\bauthor{\bsnm{Burnett}, \binits{A.D.}},
\bauthor{\bsnm{Ali}, \binits{M.}},
\bauthor{\bsnm{Sasaki}, \binits{S.}},
\bauthor{\bsnm{Linfield}, \binits{E.H.}},
\bauthor{\bsnm{Davies}, \binits{A.G.}},
\bauthor{\bsnm{Freeman}, \binits{J.R.}}:
\batitle{Effects of structural ordering on infrared active vibrations within ${\mathrm{bi}}_{2}{({\mathrm{Te}}_{(1\ensuremath{-}x)}{\mathrm{Se}}_{x})}_{3}$}.
\bjtitle{Phys. Rev. B}
\bvolume{106},
\bfpage{245203}
(\byear{2022})
\doiurl{10.1103/PhysRevB.106.245203}
\end{barticle}
\endbibitem

\bibitem[\protect\citeauthoryear{Tarakina et~al.}{2014}]{Tarakina2014}
\begin{barticle}
\bauthor{\bsnm{Tarakina}, \binits{N.V.}},
\bauthor{\bsnm{Schreyeck}, \binits{S.}},
\bauthor{\bsnm{Luysberg}, \binits{M.}},
\bauthor{\bsnm{Grauer}, \binits{S.}},
\bauthor{\bsnm{Schumacher}, \binits{C.}},
\bauthor{\bsnm{Karczewski}, \binits{G.}},
\bauthor{\bsnm{Brunner}, \binits{K.}},
\bauthor{\bsnm{Gould}, \binits{C.}},
\bauthor{\bsnm{Buhmann}, \binits{H.}},
\bauthor{\bsnm{Dunin-Borkowski}, \binits{R.E.}},
\bauthor{\bsnm{Molenkamp}, \binits{L.W.}}:
\batitle{Suppressing twin formation in bi2se3 thin films}.
\bjtitle{Advanced Materials Interfaces}
\bvolume{1}(\bissue{5}),
\bfpage{1400134}
(\byear{2014})
\doiurl{10.1002/admi.201400134}
\end{barticle}
\endbibitem

\bibitem[\protect\citeauthoryear{Egerton}{2011}]{Egerton2011}
\begin{bbook}
\bauthor{\bsnm{Egerton}, \binits{R.F.}}:
\bbtitle{Electron Energy-Loss Spectroscopy in the Electron Microscope},
\bedition{3}rd edn.
\bpublisher{Springer},
\blocation{New York}
(\byear{2011}).
\doiurl{10.1007/978-1-4419-9583-4}
\end{bbook}
\endbibitem

\bibitem[\protect\citeauthoryear{Garcia-Molina et~al.}{1985}]{GarciaMolina1985}
\begin{barticle}
\bauthor{\bsnm{Garcia-Molina}, \binits{R.}},
\bauthor{\bsnm{Gras-Martit}, \binits{A.}},
\bauthor{\bsnm{Howie}, \binits{A.}},
\bauthor{\bsnm{Ritchie}, \binits{R.H.}}:
\batitle{Retardation effects in the interaction of charged particle beams with bounded condensed media}.
\bjtitle{J. Phys. C: Solid State Phys}
\bvolume{18},
\bfpage{5335}--\blpage{5345}
(\byear{1985})
\end{barticle}
\endbibitem

\bibitem[\protect\citeauthoryear{Hage et~al.}{2013}]{Hage2013}
\begin{botherref}
\oauthor{\bsnm{Hage}, \binits{F.S.}},
\oauthor{\bsnm{Ramasse}, \binits{Q.M.}},
\oauthor{\bsnm{Kepaptsoglou}, \binits{D.M.}},
\oauthor{\bsnm{Prytz}, \binits{O.}},
\oauthor{\bsnm{Gunnaes}, \binits{A.E.}},
\oauthor{\bsnm{Helgesen}, \binits{G.}},
\oauthor{\bsnm{Brydson}, \binits{R.}}:
Topologically induced confinement of collective modes in multilayer graphene nanocones measured by momentum-resolved stem-veels.
Physical Review B
\textbf{88}
(2013)
\doiurl{10.1103/PhysRevB.88.155408}
\end{botherref}
\endbibitem

\bibitem[\protect\citeauthoryear{Smidstrup et~al.}{2019}]{Smidstrup_2020}
\begin{barticle}
\bauthor{\bsnm{Smidstrup}, \binits{S.}},
\bauthor{\bsnm{Markussen}, \binits{T.}},
\bauthor{\bsnm{Vancraeyveld}, \binits{P.}},
\bauthor{\bsnm{Wellendorff}, \binits{J.}},
\bauthor{\bsnm{Schneider}, \binits{J.}},
\bauthor{\bsnm{Gunst}, \binits{T.}},
\bauthor{\bsnm{Verstichel}, \binits{B.}},
\bauthor{\bsnm{Stradi}, \binits{D.}},
\bauthor{\bsnm{Khomyakov}, \binits{P.A.}},
\bauthor{\bsnm{Vej-Hansen}, \binits{U.G.}},
\bauthor{\bsnm{Lee}, \binits{M.-E.}},
\bauthor{\bsnm{Chill}, \binits{S.T.}},
\bauthor{\bsnm{Rasmussen}, \binits{F.}},
\bauthor{\bsnm{Penazzi}, \binits{G.}},
\bauthor{\bsnm{Corsetti}, \binits{F.}},
\bauthor{\bsnm{Ojanperä}, \binits{A.}},
\bauthor{\bsnm{Jensen}, \binits{K.}},
\bauthor{\bsnm{Palsgaard}, \binits{M.L.N.}},
\bauthor{\bsnm{Martinez}, \binits{U.}},
\bauthor{\bsnm{Blom}, \binits{A.}},
\bauthor{\bsnm{Brandbyge}, \binits{M.}},
\bauthor{\bsnm{Stokbro}, \binits{K.}}:
\batitle{Quantumatk: an integrated platform of electronic and atomic-scale modelling tools}.
\bjtitle{Journal of Physics: Condensed Matter}
\bvolume{32}(\bissue{1}),
\bfpage{015901}
(\byear{2019})
\doiurl{10.1088/1361-648X/ab4007}
\end{barticle}
\endbibitem

\bibitem[\protect\citeauthoryear{{van Setten} et~al.}{2018}]{VANSETTEN201839}
\begin{barticle}
\bauthor{\bsnm{{van Setten}}, \binits{M.J.}},
\bauthor{\bsnm{Giantomassi}, \binits{M.}},
\bauthor{\bsnm{Bousquet}, \binits{E.}},
\bauthor{\bsnm{Verstraete}, \binits{M.J.}},
\bauthor{\bsnm{Hamann}, \binits{D.R.}},
\bauthor{\bsnm{Gonze}, \binits{X.}},
\bauthor{\bsnm{Rignanese}, \binits{G.-M.}}:
\batitle{The pseudodojo: Training and grading a 85 element optimized norm-conserving pseudopotential table}.
\bjtitle{Computer Physics Communications}
\bvolume{226},
\bfpage{39}--\blpage{54}
(\byear{2018})
\doiurl{10.1016/j.cpc.2018.01.012}
\end{barticle}
\endbibitem

\bibitem[\protect\citeauthoryear{Monkhorst and Pack}{1976}]{Monkhorst1976}
\begin{barticle}
\bauthor{\bsnm{Monkhorst}, \binits{H.J.}},
\bauthor{\bsnm{Pack}, \binits{J.D.}}:
\batitle{Special points for brillouin-zone integrations}.
\bjtitle{Phys. Rev. B}
\bvolume{13},
\bfpage{5188}--\blpage{5192}
(\byear{1976})
\doiurl{10.1103/PhysRevB.13.5188}
\end{barticle}
\endbibitem

\bibitem[\protect\citeauthoryear{Grimme et~al.}{2010}]{Grimme2010}
\begin{barticle}
\bauthor{\bsnm{Grimme}, \binits{S.}},
\bauthor{\bsnm{Antony}, \binits{J.}},
\bauthor{\bsnm{Ehrlich}, \binits{S.}},
\bauthor{\bsnm{Krieg}, \binits{H.}}:
\batitle{{A consistent and accurate ab initio parametrization of density functional dispersion correction (DFT-D) for the 94 elements H-Pu}}.
\bjtitle{The Journal of Chemical Physics}
\bvolume{132}(\bissue{15}),
\bfpage{154104}
(\byear{2010})
\doiurl{10.1063/1.3382344}
\end{barticle}
\endbibitem

\bibitem[\protect\citeauthoryear{Neugebauer and Scheffler}{1992}]{Neugebauer1992}
\begin{barticle}
\bauthor{\bsnm{Neugebauer}, \binits{J.}},
\bauthor{\bsnm{Scheffler}, \binits{M.}}:
\batitle{Adsorbate-substrate and adsorbate-adsorbate interactions of na and k adlayers on al(111)}.
\bjtitle{Phys. Rev. B}
\bvolume{46},
\bfpage{16067}--\blpage{16080}
(\byear{1992})
\doiurl{10.1103/PhysRevB.46.16067}
\end{barticle}
\endbibitem

\bibitem[\protect\citeauthoryear{Makov and Payne}{1995}]{Makov1995}
\begin{barticle}
\bauthor{\bsnm{Makov}, \binits{G.}},
\bauthor{\bsnm{Payne}, \binits{M.C.}}:
\batitle{Periodic boundary conditions in ab initio calculations}.
\bjtitle{Phys. Rev. B}
\bvolume{51},
\bfpage{4014}--\blpage{4022}
(\byear{1995})
\doiurl{10.1103/PhysRevB.51.4014}
\end{barticle}
\endbibitem

\end{thebibliography}

\end{document}